\documentclass[12pt]{iopart}
\usepackage{iopams}
\usepackage{graphicx}
\usepackage{csquotes}
\usepackage{microtype}
\usepackage{url}

\usepackage[numbers,sort&compress]{natbib}
\usepackage{hyperref}
\usepackage{doi}\usepackage{etoolbox}
\usepackage{siunitx}

\newcommand\elementcharge[2]{{\expandafter\MakeUppercase #1}\textsuperscript{#2}}
\newcommand\isotope[2]{\textsuperscript{#2}{\expandafter\MakeUppercase #1}}
\newcommand\isotopecharge[3]{\textsuperscript{#2}{\expandafter\MakeUppercase #1}\textsuperscript{#3}}
\newcommand\moleculemass[3]{\textsuperscript{#2}{\expandafter\MakeUppercase #1}\textsubscript{#3}}

\begin{document}
\title[Tutorial laser locking techniques and vapor cells]{Tutorial on laser locking techniques and the manufacturing of vapor cells for spectroscopy}
\author{Max M\"ausezahl$^1$, Fabian Munkes$^1$, and Robert L\"ow$^1$}
\address{$^1$ 5.~Physikalisches Institut and Center for Integrated Quantum Science and Technology,
	Universit\"at Stuttgart, Pfaffenwaldring 57, 70569 Stuttgart, Germany}
\ead{r.loew@physik.uni-stuttgart.de}
\vspace{10pt}
\begin{indented}
\item[]Januar 2024
\end{indented}
\begin{abstract}
This tutorial provides a hands-on entry point about laser locking for atomic vapor
research and related research such as laser cooling.
We furthermore introduce common materials and methods for the fabrication of vapor cells
as a tool for this research.
Its aim is not to be exhaustive, but rather to provide an overview about the
possible techniques that are actively employed in labs today.
Some critical parameters of
locked laser system for use with thermal atomic vapors are introduced and discussed.
To exemplify this, we describe a versatile locking system that caters for many of the needs we found during
our research with thermal atomic vapors.
We also emphasize the compromises we took during our decision-making process.
\end{abstract}
\vspace{2pc}
\submitto{\NJP}
\section{Introduction}
Quantum optical experiments with thermal or cooled atomic vapors require controlled access to the
electronic level structure of the atomic system.
Since the interaction between these electronic states emits or absorbs electromagnetic
radiation, this often means that precisely tuned lasers are employed to prepare and probe atomic
ensembles.
The toy example of a two level, dipole allowed atomic transition between a ground state $|g\rangle$
and an excited state $|e\rangle$ can be used to derive some principle requirements for such a
radiation source:
It needs to match the frequency  $\nu_0=|E_{|g\rangle}-E_{|e\rangle}|/h$ with a linewidth ideally
smaller than the linewidth $\Delta\nu$ of the atomic transition.
This ensures that the atoms effectively couple to the electromagnetic field.
The usage of long-lived clock transitions as used in atomic clocks means that
$\Delta\nu/\nu_0$ can be as small as \num{2.5e-19}~\cite{Marti.2018}, which is a tough
requirement for the laser stability in the presence of external influences~\cite{Matei.2017}.
\par
The specific case of thermal vapors sometimes relaxes the requirements regarding the precision
of the absolute laser frequency $\nu$ and its linewidth $\Delta\nu$ in
comparison to experiments with cold atoms.
This is due to the Doppler-shift and associated Doppler-broadening resulting from the thermal
Maxwell-Boltzmann velocity distribution.
Within the Doppler-width $\Delta_\mathrm{D}$, a significant number of atoms can be expected to fall
in a velocity class $\bi{v}$ that compensates any residual detuning
$\Delta=\bi{k_\mathrm{l}}\cdot \bi{v}$ (laser wave vector $\bi{k_\mathrm{l}}$).
Note however, that this is only true as long as only a single transition is investigated.
In multi level excitation schemes, a sub-Doppler selection based on the interplay
between the individual transitions is possible.
\par
In either case the so-called free-running stability of available laser systems is usually not
sufficient for experiments with atomic vapors.
They suffer from a non-constant center frequency $\nu$ movement called \enquote{drift}
due to varying environmental conditions or aging.
Additionally, the arising laser linewidth $\Delta\nu$ is broadened due to internal noise
processes or limited spectral selectivity of the laser cavity.
Both issues can be solved by \enquote{laser locking} which is a coined phrase for providing
feedback to the laser in a feedback loop to stabilize its frequency (sometime also called
\enquote{laser stabilization} or \enquote{frequency stabilization}).
A reduction of laser linewidth is usually referred to as \enquote{linewidth narrowing}.
\par
Laser locking is a tool commonly employed by many laboratories working with lasers
in the fields of atomic physics, molecular physics, or optical metrology.
Commercial turn-key solutions for many applications are readily available.
Nevertheless, the technology is still subject of current research on a pursuit towards smaller
packaging~\cite{Groswasser.2009, Lee.2011,
Dinkelaker.2017, Maurice.2020, Lee.2021}, even narrower linewidths~\cite{Salomon.1988, Alnis.2008,
Kessler.2012, Torrance.2016, Matei.2017}, better long-term stability~\cite{Lee.2020} and usability
features like  AI support~\cite{Li.2020}, autonomous or automatic operation~\cite{Koch.2003,
Allard.2004, Canuto.2007, Sparkes.2011, Barwood.2012, Dinkelaker.2017, Roy.2019, Li.2020, Guo.2022} and
cost-efficiency~\cite{Yang.2012, Huang.2014, Leibrandt.2015, Jrgensen.2016,
Subhankar.2019, Roy.2019, Preuschoff.2020, Strangfeld.2022, Neuhaus.2023}.
This tutorial neither aims to be an exhaustive review of all possible locking schemes with their
respective advantages, nor to roll out the full underlying theory of feedback loops in general
or their specific application with laser systems.
It is much rather meant as a practical guide with a brief theoretical foundation for the undergraduate
or graduate student while being exposed to locked lasers for the first time.
It is also targeted towards senior researchers, who are planning experiments with hot atomic
vapors and looking into the manifold options that exist today ---
possibly based on an existing variety of laser systems and controllers, which they wish to upgrade.
\par
Introducing or getting introduced to laser locking in an accurate and adequately formal way without
getting lost in the many details is not any easy task and can feel overwhelming.
It is not a narrow subject and much rather sits at the interface between control theory,
radio frequency (RF) operation, and optics.
Each of those subjects is worth dozens of textbooks on its own.
Luckily, a wide range of well-documented solutions exist, and it is certainly possible to
set up intricate experiments without dedicating years of learning.
\par
We purposely suggest a larger number of references,
which could serve as a starting point for additional reading on the individual methods in
the various sections.
Moreover, there are a number of existing review articles and books covering the topic of laser
locking.
Some of them are technically mostly outdated due to their age but still feature a solid
introduction into the fundamental principles~\cite{White.1965, Hall.1968, Baird.1974, Hall.1978,
Ohtsu.1988, Hamilton.1989, Roberts.1999}.
Recent comprehensive publications on the matter include the articles by
Hall~et~al.~\cite{Hall.2010}, by Taubman~\cite{Taubman.2013}, and by
Wu~et~al.~\cite{Wu.2021} as well as the respective chapter in a book by
Fang~et~al.~\cite{Fang.2017}.
Additionally, there are in-depth publications covering a specific set or aspect of laser
locking techniques~\cite{Fox.2003, Gawlik.2004}.
\section{Characterizing laser frequency fluctuations}
Previously, we introduced the ad-hoc terms of external noise, laser linewidth, and center
frequency drift.
As physicists, we might already have an intuitive sense of the meaning and delicate interplay
between these quantities, e.g. how stronger acoustic noise could lead to a larger laser linewidth.
A rigorous treatment and precise definition of these details is possible by observing
frequency domain spectra or time domain variances \cite{Sullivan.1990,Zhu.1993}.
There are numerous standardized quantities which can be explored \cite{Riley.2008},
of which we will introduce the widely used Allan variance in time \cite{Allan.1966}, and
power spectral density (PSD) in frequency \cite{Roberts.1999}.
\par
We start from the oscillating, instantaneous optical field amplitude $E(t)$ (or other source
amplitude), which is typically not directly accessible in an optics experiment.
This field has an instantaneous frequency $f(t)$ and a total, average power
$P=\alpha\lim_{T\to\infty}\int_{-\infty}^{\infty} |E(t)|^2 dt$ (with some factor $\alpha$ depending
on the type of the field).
\par
The power spectral density $P_E(f)$ of this field $E$ is then defined using the Fourier transform of the
autocorrelation according to the Wiener-Khinchin theorem, such that the total power within a
certain frequency interval $f_1\leq f \leq f_2$
is given by the integral over this frequency range (again with some normalization factor $\beta$):
\begin{eqnarray}
P_{[f_1,f_2]}=\beta\int_{f_1}^{f_2} P_E(f) df.
\end{eqnarray}
In practice, it is this integral formula that makes the power spectral density useful to understand short-term
effects of fractional power located close to the desired optical target frequency.
Since the power spectral density of lasers used in atomic physics is usually narrowband, a full width at half maximum (FWHM)
linewidth of this power spectral density is often given in literature as a single figure of performance.
The aforementioned drift of the laser frequency describes a non-constant peak frequency of the
power spectral density when repeating the measurement over time.
The power spectral density can be directly optically measured using a scanning analysis resonator,
an autocorrelator, or a beat-note analysis incorporating a second reference laser
\cite{Roberts.1999}.
\par
The Allan variance on the other hand is understood as a measure of the long-term performance of
an oscillator, i.e. the long-term stability of the laser frequency.
Like the standard variance it measures a statistical spread of a set of numbers (in this case
frequency differences or phase differences) but does not diverge for common noise sources such
as flicker noise.
A concrete measurements captures $N$ frequency deviations $\Delta f_n$ with $n\in\{1,2,...,N\}$
between two laser systems from a beat-note signal averaged over a (short) observation times $\tau_0$ a
certain times $t_n=n\tau_0$ (we assume zero dead time between measurements here).
These are often converted to dimensionless fractional frequencies $y_n=\Delta f_n / f_0 -1$.
The Allan variance $\sigma_y^2(\tau)$ is then calculated as
\begin{eqnarray}
\sigma_y^2(\tau) = \frac{1}{2 (N-1)} \sum_{i=0}^{N-1} (y_{i+1} - y_{i})^2.
\end{eqnarray}
This can be intuitively understood like a standard variance --- notice the general form of an
time average of squared deviations --- that however takes into consideration deviations between
successive samples instead of deviations from the overall mean.
Conventionally one plots the square root of the Allan variance, called Allan deviation
$\sigma_y(\tau)$, on a log-log plot over the total measurement time $\tau$.
Such plots are exceptionally rich to the experienced eye \cite{Sullivan.1990}, so that we only
give two brief starting points for interpretation here:
A small value of $\sigma_y(\tau)$ at a certain time $\tau$ indicates a small root-mean-square
frequency deviation between the two lasers at over this time span.
The slopes appearing in such a plot are an indicator for the underlying type of noise process,
which follow specific power laws.
\par
Today, use of the original Allan variance has been largely superseded by extended and
improved variants like the overlapping Allan variance or the modified Allan variance
\cite{Riley.2008}.
\section{The laser feedback loop}
It is impossible to talk about laser locking without a basic understanding of feedback  loops and
control theory.
Control loops are hidden in many of our modern devices, e.g.\ in each switched-mode
USB power supply in order to deliver a precise voltage to mobile devices.
They are even more common to experimental physicists, who are dealing with a plethora of
heterogeneous systems, which need to be held within tight operational windows,
such as a vapor cell that is kept at a specific temperature by a thermal controller.
This section introduces their most important concepts in the context of laser locking specific
issues.
A broader and more detailed introduction into \enquote{feedback for physicists} can be found in
the identically named reference~\cite{Bechhoefer.2005}.
\par
The first step to understand, how the frequency of a laser can be stabilized, is to recognize,
why this becomes necessary at all.
A textbook laser consists of an optical resonator (called cavity here), an active medium,
and an energy pump, which define the emission spectrum.
In reality, there are environmental influences that result in a non-constant laser frequency
over time owing to temperature, humidity and air pressure fluctuations, mechanical vibrations,
aging effects, and even external electromagnetic fields.
These lead to varying refractive indices, changing lengths, and emission
properties making the overall system highly unpredictable.
Therefore, a \enquote{passive stabilization} of the laser system is mandatory even when active laser
locking is employed.
As reference~\cite{Talvitie.1997} puts it: \enquote{Generally, the first step to improve the laser
performance is to make the cavity construction insensitive to these external effects.}
It must be stressed that it might be impossible to reliably lock a laser, if the overall
mechanical structure is too unstable.
\par
\begin{figure}
	\centering
	\includegraphics{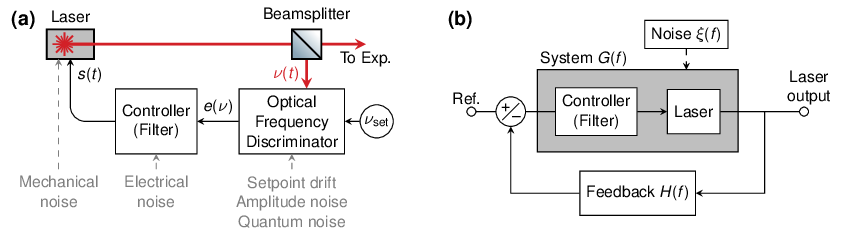}
	\caption{
		Two schematic representations of a laser locking feedback loop, both adapted from
		reference~\cite{Hall.2010}.
		(a) A model showing the key components of the system, starting with a laser that emits a
		certain	noise-disturbed and time-dependent frequency $\nu(t)$, of which
		a portion is split off and sent to an optical frequency discriminator.
		This compares it with some setpoint frequency $\nu_\mathrm{set}$ and produces an error signal
		$e(\nu)$ that is processed by a controller to counteract any disturbances via
		control inputs $s(t)$ of the laser.
		(b) A schematic view of the same system from a control theory standpoint, where the action of
		each component is represented by a complex valued, frequency $f$ domain transfer function.
		$H(f)$ describes the amplitude response and phase-lag of the feedback sensor system,
		while $G(f)$ represents both the controller and the laser.
	}
	\label{fig:controller}
\end{figure}

There is however a point, at which it becomes technically infeasible to add more passive stability,
as this usually comes with a larger build volume and an excessively increasing cost.
In such cases, the initially \enquote{open-loop} system is expanded by a feedback loop, which
monitors the actual, instantaneous laser frequency $\nu(t)$ through an optical frequency
discriminator (\fref{fig:controller}(a)).
The stability issue is then transferred to a comparison between $\nu(t)$ and a known stable
frequency  reference $\nu_\mathrm{set}$.
Many optical frequency discrimination devices have been developed, some of which are covered in
\sref{sec:commonerrorsignaltechniques}.
\par
In each case the optical frequency discriminator produces an error signal $e(\nu)$, which
carries the information about momentary state of $\nu(t)$ compared to $\nu_\mathrm{set}$.
This signal must thus contain both the sign (i.e. $\nu<\nu_\mathrm{set}$ or
$\nu>\nu_\mathrm{set}$) and the magnitude $|\nu-\nu_\mathrm{set}|$ of any deviation.
It is therefore often encoded as a voltage signal roughly proportional to
$\nu-\nu_\mathrm{set}$.
This ideal case is rarely achieved, but a decent approximation is sufficient for a suitable
controller (also called filter due to its inner working principle).
By setting a number of control signals $s(t)$ of the laser, the controller counteracts the
deviation as reported by the optical frequency discriminator.
\Fref{fig:controller}(a) also includes various sources of noise and disturbances which are
imperfections of the system.
It is ultimately the goal of the operator to tune the various filters and signal amplitudes
such that external noise sources are effectively suppressed, without introducing additional noise.
\par
Reasoning about the properties of the overall system in the time domain as a control theory problem
is usually impractical and thus much rather performed in the frequency domain $f$ of all
noise sources $\xi(f)$~\cite{Hall.2010}.
In this representation (\fref{fig:controller}(b)), controller and laser form a frequency dependent
system $G(\nu)$ which responds to feedback $H(f)$ in the presence of certain noise
$\xi(f)$.
Notice, that $G$ and $H$ generally are complex, normalized transfer functions which include
phase-lag of the various components, and that the frequency $f$ is not an optical frequency
close to $\nu$,
but a much smaller, unwanted oscillation frequency component disturbing the operation.
The new overall system has a transfer function~\cite{Hall.2010}
\begin{eqnarray}
	L(f)=\frac{G(f)}{1+G(f)H(f)}, \label{eq:transferfunction}
\end{eqnarray}
which can be tuned through the controller settings and the feedback sensors.
On a first glance, making the amplitude of $H$ as large as possible would in theory
eliminate all noise, however a closer look identifies time lag (and therefore phase lag)
of the feedback $H(f)$ as the ultimate limit.
For $|GH|=-1$ self-reinforcing positive feedback (out of phase by
\SI{180}{\degree}, also called feed-forward) is achieved.
In these circumstances the laser emission frequency would start to oscillate (modulate) at such
frequencies on its own, which is the contrary of a stabilized laser.
Therefore, feedback loops are carefully designed and then hand-tuned to strongly counteract
unwanted frequency components, e.g. acoustical frequencies in the \si{\kilo\hertz} range
with high gain, while the inevitable frequency, at which the phase lag becomes larger
than \SI{180}{\degree} is appropriately dampened.
It follows, that each feedback loop has a usable feedback
bandwidth $f_\mathrm{max}$ above which frequency noise can not be sufficiently reduced.
As a rule of thumb (assuming a reasonable choice of the control parameters) a high bandwidth is
necessary to reduce high-frequency noise and narrow the laser linewidth.
Small feedback bandwidths could nevertheless be sufficient to eliminate drifts.
\par
A basic setup might use a proportional-integral-derivative (PID) controller in combination with
various low-pass filters to suppress these resonance points:
A proportional controller that directly counteracts $e(\nu)$, an integral part that
corrects any residual error $\langle e(\nu)\rangle_t$ (\enquote{proportional droop} at
low frequencies~\cite{Bechhoefer.2005}), and a derivative part that operates at high frequencies,
trying to \enquote{predict} rapid changes of the system e.g.\ due to a singular external
disturbance.
Modern commercial controllers with high bandwidth offer a wide range of settings such that
\enquote{optimal control} (a coined term for the \enquote{best} possible settings) might only be
possible by employing computer simulations and rigorous measurements.
Controllers can also include usability features like automatic tuning~\cite{Ziegler.1993} or
automatic locking if the frequency is lost (re-locking).
A common manual procedure to lock a so far unknown laser, called \enquote{tuning} the controller,
is presented in \sref{sec:oursetup}.
\par
Beyond the (electrical) controller itself, the response of the laser to its control parameters $s$
including all delays is described within $G(f)$ in \eref{eq:transferfunction}.
Consequently, a high-bandwidth controller requires a high-bandwidth response of the
laser itself regarding the control signals $s(t)$.
Laser spectroscopy with atoms and molecules sees common use of various laser types that have
different characteristics in this regard:
\begin{itemize}
	\item External-cavity diode lasers, sometimes also called extended-cavity diode lasers, (ECDLs)
	are commercially available with free-running linewidths of few \SI{100}{\kilo\hertz} and
	tuning over tens of \si{\nano\meter}~\cite{Saliba.2009}.
	High bandwidth control (many \si{\mega\hertz}) is possible through the pump current which has
	the unwanted side effect of output power modulation.
	Therefore, it is usually combined with the mechanically limited, lower bandwidth piezoelectric
	length tuning of the external cavity (few \si{\kilo\hertz}).
	Low bandwidth control based on diode temperature is mostly seen as a part of the passive
	stability of the system, i.e.\ it is not part of the laser frequency feedback loop and rather
	controlled by a different controller using a dedicated temperature
	sensor~\cite{Ricci.1995,Fleming.1981}.
	\item Distributed-feedback diode lasers (DFBs) and the closely related distributed Bragg resonator lasers (DBRs)
	are similar but mechanically simplified and thus cost reduced alternatives
	to ECDLs with less tuning
	range (few \si{\nano\meter}) and larger free-running linewidths, that can nevertheless be
	greatly reduced through laser locking and the accompanying linewidth
	narrowing~\cite{Bennetts.2014}.
	Again, high bandwidth control (many \si{\mega\hertz}) is available through the pump current
	while large wavelength changes and low-frequency control (few \si{\hertz}) is provided through
	the chip temperature.
	\item Fiber based single-frequency lasers are typically either a DFB or
	a DBR implemented inside a fiber and thus feature the same control mechanisms~\cite{Shi.2014}.
	\item Dye lasers have historically been the first widely tuneable, high-power lasers systems
	available at a wide range of wavelengths that could be
	simultaneously narrowed to linewidths below \SI{1}{\mega\hertz} \cite{Hansch.1973,Duarte.1990}.
	Coarse tuning is performed using cavity length tuning, which can be part of the feedback loop.
	Finer mode-hop free tunability is then typically achieved by placing a dispersive, piezoelectric
	actuator driven	(several \SI{10}{\kilo\hertz} feedback bandwidth)
	wavelength selector such as a Littrow configuration grating, a Fabry-Pérot interferometer,
	or a birefringent filter such as a Lyot filter inside the laser resonator\cite{Walther.1974}.
	Higher feedback bandwidth can be achieved using acousto-optic or electro-optic modulators
	inside the cavity \cite{Hollberg.1990}.
	\item Solid state titanium-sapphire lasers (Ti:sas) are currently the predominant
	femtosecond laser source due to their large spectral
	bandwidth and tuning range. They are built similar to
	dye lasers yet overcome many of their detrimental properties and can be
	tuned by intra-cavity elements just as described above \cite{Wall.1990,Duarte.2016}.
	\item Vertical-external-cavity surface-emitting lasers (VECSELs) have recently seen increased use in the quantum optics community as they
	bridge a gap between the low-power but easily tuned semiconductor lasers and the high-power
	solid state lasers \cite{Guina.2017}.
	While electrically pumped VECSELs exist, 	optically pumped VECSELs currently have
	a technological edge.
	Thus, the tuning mechanisms in available VECSELs systems are comparable to dye
	systems and Ti:sas using cavity tuning and intra-cavity elements \cite{Burd.2023}.
\end{itemize}
Finally, it is possible to add an arbitrary, posterior high-bandwidth control to any laser by adding
an optical frequency modulator to the output, which can then serve as a control point in the
feedback loop \cite{Hall.1978}.
\section{RF laser modulation techniques}
We write the temporally changing electrical field vector $\bi{E}(t)$ of a
linearly polarized, monochromatic plane wave at a fixed position as
\begin{eqnarray}
	\bi{E}(t) &= \bi{e}(t)E_0(t)\exp \left[ i\left(2\pi \int_{0}^{t}\nu(\tau) d\tau + \varphi(t)\right) \right] \label{eq:monowave} \\
	&= \bi{e}(t)E_0(t)\exp \left[ i\left(2\pi \nu t + \varphi(t)\right) \right]. \label{eq:monowaveeasy}
\end{eqnarray}
Modulation refers to the periodic variation of either the polarization unit vector $\bi{e}(t)$,
the amplitude $E_0(t)$, the frequency $\nu(t)$, or the phase $\varphi(t)$ with a
certain modulation frequency $f_\mathrm{mod}$.
Notice that \eref{eq:monowaveeasy}, which is commonly found in textbooks, is only true
in case of constant frequency $\nu$.
Usage of optical RF modulation within the context of
laser locking falls broadly into one of three applications:
\begin{enumerate}
	\item Frequency sideband generation as required by many frequency discrimination techniques,
	\item offsetting, postlaser-correcting \cite{Ma.1994}, or sweeping an already locked laser
		frequency, or
	\item adding an arbitrary, posterior, high-bandwidth control within the feedback loop, when it is
		not easily available otherwise \cite{Hall.1977, Hall.1978}.
\end{enumerate}
Acousto-optic modulators (AOMs) directly shift the frequency of the diffracted light by some multiple of the
applied acoustic wave frequency and can thus be used as frequency modulators or frequency shifters
with sub \si{\giga\hertz} bandwidth.
Electro-optic modulators (EOMs) exists as phase, polarization, and amplitude modulators, of which phase modulators
with multi \si{\giga\hertz} bandwidth are particularly common to laser locking.
\par
Notice that, for most practical sideband generation purposes, frequency modulation (FM) and
phase modulation (PM) can be used interchangeably, as they are specific cases of angle modulation.
Therefore, an additional modulator might be omitted in favor of direct laser current modulation
which leads to FM.
The latter comes at the price of frequency sidebands present in the main beam,
which is sometimes acceptable, e.g.\ if the main beam is subsequently filtered through a cavity
as part of a second harmonic generation process, or if it solely serves as a reference for other
lasers.
Nevertheless, AOMs or EOMs will usually be found in separate side beams dedicated to
laser locking, while the main beam stays free of sideband modulation and is controlled by means
of any other aforementioned control signal.
\begin{figure}
	\centering
	\includegraphics{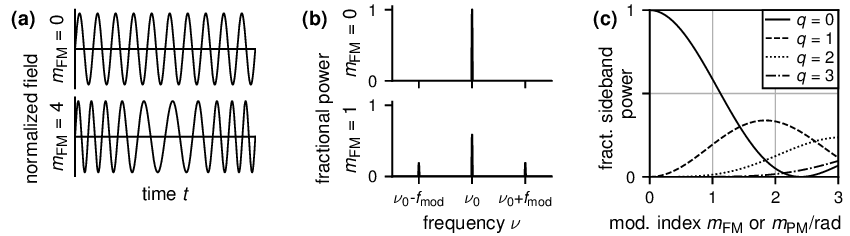}
	\caption{
		(a) The temporal shape of an unmodulated, normalized, sinusoidal electric field amplitude in
		comparison to an FM sinusoidal signal with the same carrier frequency $\nu$ and
		$f_\mathrm{mod} = \nu/10$.
		(b) The spectral power of an unmodulated, sinusoidal signal and an FM signal normalized to
		the total power.
		(c) The spectral power per sideband over different modulation indices for the carrier and the
		lowest three sidebands. Throughout the plotted region the combined power in
		all other sidebands $q>3$ is less than \SI{4}{\percent}.
	}
	\label{fig:RF_modulation}
\end{figure}
\par
Sinusoidal FM or PM in \eref{eq:monowave} are given by
\begin{eqnarray}
	\nu(t)&=\nu_0+\hat{\nu}_\mathrm{mod}\cos\left(2\pi f_\mathrm{mod}t\right), \qquad \text{with } 0 \leq \hat{\nu}_\mathrm{mod} \leq \nu_0 \text{ and} , \label{eq:PM} \\
	\varphi(t)&=\varphi_0+\hat{\varphi}_\mathrm{mod}\sin\left(2\pi f_\mathrm{mod}t\right) \label{eq:FM}
\end{eqnarray}
respectively.
In both cases the modulation leads to the creation of sidebands at multiples of $f_\mathrm{mod}$,
i.e. at $\nu_0 \pm qf_\mathrm{mod}, q\in\left\{0,1,2,...\right\}$.
Even for a perfectly sinusoidal modulation all those sidebands are present
with some nonzero amplitude (see \fref{fig:RF_modulation}(a) and (b)).
It turns out that the spectral power which is transferred into each
sideband is conveniently described by a modulation index $m$ which depends on the peak frequency
or phase variation and the modulation frequency:
\begin{eqnarray}
	m_\mathrm{FM}&=\hat{\nu}_\mathrm{mod}/f_\mathrm{mod}\qquad&\text{(dimensionless)}, \\
	m_\mathrm{PM}&=\hat{\varphi}_\mathrm{mod}&\text{(radians)}.
\end{eqnarray}
Note, that while we follow the convention of reference \cite{Roder.1931},
the term modulation index is sometimes used interchangeably with the term modulation depth
\cite{Harris.1965, Black.2001}.
The precise amplitude of each sideband of order $q$ is given by Bessel
functions of first kind as $J_q(m_\mathrm{FM})$ or $J_q(m_\mathrm{PM}/\si{\radian})$ \cite{Roder.1931}.
Squaring these amplitudes results in the sideband spectral power as plotted in
\fref{fig:RF_modulation}(c).
As there are always two symmetrical sidebands except for the carrier at $q=0$, the total power is
conserved due to $J_0^2(x)+2\sum_{q=1}^{\infty}J_q^2(x)=1$ (this also means that
phase modulation does not add power to the field) \cite{Olver.2010}.
Finally, since any arbitrary periodic signal can be understood as a linear Fourier composition of
multiple sinusoidal terms, the presented relations can be transferred accordingly.
\section{Overview of common frequency discrimination techniques}
\label{sec:commonerrorsignaltechniques}
While the selection of a suitable electronic controller as discussed in the previous sections is
undoubtedly important for the eventual success of the laser lock, there are a number of commercial
and open-source solutions that work interchangeably for a wide range of applications.
This section now describes the optical part of the laser lock, which is typically
a deliberate choice based on a case-by-case evaluation.
In this tutorial we try to provide an overview and classification of various commonly deployed
optical frequency discrimination techniques.
This selection as outlined in \tref{tab:methods} is based on appearance in recent publications but also
includes some less common variations mainly for their historic significance or educational value
in the overall context.
\begin{table}
	\centering
	\caption{\label{tab:methods}
		Overview over most frequency discrimination methods discussed in the individual subsections in
		the main text based on the information given in the references of these descriptions.
		The \enquote{reference} column indicates which component or oscillator serves as a frequency
		reference.
		Atomic references are absolute, while cavities and interferometers need to be calibrated with
		an external standard.
		The \enquote{RF} column lists, whether RF modulation and demodulation of laser light
		is required.
		Typical bandwidths have been extracted from the various references and can only give a
		rough indication, while specific modifications might allow or require much higher or lower
		values.
		The last column shows, whether the technique is of interferometric nature, i.e.\ whether it
		is based on the coherent interaction between multiple fields on the detector.
	}
	\footnotesize
	\lineup
	\begin{tabular}{@{}lcccc}
		\br{}
		&&&Typical band-& Interfer-\\
		Method&Reference&RF&widths (\si{\kilo\hertz})&ometric\\
		\mr{}
		Side of fringe&Various&no& \0\0$>10$ &no\\
		Dither lock&Various&no&\0$>100$ &no\\
		\mr{}
		Pound-Drever-Hall (PDH)&Cavity&yes&$>1000$&yes\\
		Hänsch-Couillaud&Cavity&no&\0$>100$ &yes\\
		\mr{}
		Saturated absorption spectroscopy (SAS)&Atomic&no&\0\0$>10$ &no\\
		Polarization spectroscopy&Atomic&no&\0$>100$ &no\\
		Dichroic atomic vapor laser lock (DAVLL)&Atomic&no&\0$>100$ &no\\
		Frequency modulation spectroscopy (FMS)&Atomic&yes& $>1000$ &yes\\
		Modulation transfer spectroscopy (MTS)&Atomic&yes&\0$>100$ &yes\\
		\mr{}
		(Scanning) transfer cavity &Laser&& varying &\\
		Optical frequency lock loop (OFLL)&Laser&yes& \0$>100$&yes\\
		Optical phase lock loop (OPLL)&Laser&yes& \0\0$>10$ &yes\\
		Optical frequency comb (OFC)&Laser&yes& $>1000$ &yes\\
		\mr{}
		Interferometer / Wavemeter&Various&no& \0\0\0$<1$ &yes\\
		\br{}
	\end{tabular}
\end{table}
\par
The central goal of each optical frequency discriminator is the creation of a typically dispersive
error signal $e(\nu)$ with respect to a target laser frequency $\nu_\mathrm{set}$ that can
be fed into an electronic controller.
This is done by comparing the actual laser frequency with a known absolute or relative frequency
reference.
Absolute references include atomic and molecular transitions which lend themselves particularly
well as a source of stability when working with their vapors and form the foundation of modern
time standards~\cite{Marti.2018}.
Optical cavities, interferometers, and reference lasers are relative references, that will drift
over time if exposed to environmental changes, but can be stabilized and calibrated to serve as a
fixed, almost absolute frequency if necessary.
\par
Another classification of each technique can be made based on the frequency
range $[\nu_\mathrm{min}, \nu_\mathrm{max}]$, where the sign of the error signal properly matches
the actual sign of $\nu$ with respect to $\nu_\mathrm{set}$.
This so-called capture range is usually limited due to bandwidth caps or periodic signal features.
Additionally, each technique has a maximum usable bandwidth, either due to inherent limitations
such as a modulation frequency, due to sensor bandwidth, or due to a finite sampling rate.
For many techniques RF modulation and demodulation of the laser light is required.
While this usually leads to a reduction of noise at the modulation sidebands comparable to a
lock-in amplifier, it also comes at the expense of additional AOMs, EOMs, and RF
hardware and usually limits the bandwidth to the chosen modulation frequency.
Bandwidth alone is yet not sufficient, as there must be strong frequency discrimination (usually
expressed as a slope of $e(\nu)$ around the target frequency).
If the spectral reference feature is broad, this usually leads only to weak response for
small frequency changes compared to the overall electric noise in the system and therefore
a bad signal-to-noise ratio (SNR).
Finally, it is possible to classify each scheme based on the presence of interferometric coherence
for signal generation: Some solely depend on a single, incoherent intensity measurement, while others
interfere multiple fields coherently~\cite{Harvey.2003}.
\subsection{Locking to an optical cavity}
\label{sec:cavitylock}
\begin{figure}
	\centering
	\includegraphics{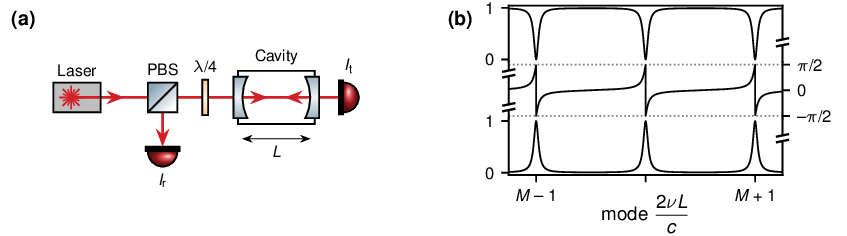}
	\caption{
		A linear optical cavity (drawing and plot adapted from references~\cite{Taubman.2013, Siegman.1986}).
		(a) Schematic drawing of the cavity with length $L$ including photodetectors for the
		transmitted intensity $I_\mathrm{t}$ and the reflected intensity $I_\mathrm{r}$.
		Assumes a linearly polarized input beam such that the ideal cavity and a
	  quarter wave plate aligned under \SI{45}{\degree} to the polarization axis will lead to the
		reflected beam being also reflected at the polarizing beam splitter (PBS).
		(b) From top to bottom: The reflected intensity, phase shift of the reflected light field
		$\Delta\varphi_\mathrm{r}$, and the transmitted intensity normalized to
		$I_\mathrm{t}+I_\mathrm{r}$  at a finesse $\mathcal{F}=20$.
	}
	\label{fig:cavities}
\end{figure}
An optical cavity or resonator like shown in \fref{fig:cavities}(a) is
formed by two (or more) mirrors, between which a resonant electromagnetic field can build up.
In this example a quarter wave plate is added in front of the cavity such that the incident
linearly polarized beam is
converted to a circular polarization and then, after reflection, again to a linearly polarized
beam with a \SI{90}{\degree} rotated polarization plane.
This clearly separates the reflected beam from the incident beam.
We assume no energy loss due to absorption in the following discussion for simplicity,
such that the transmitted intensity $I_\mathrm{t}$ and the reflected intensity $I_\mathrm{r}$ sum up
to the incident intensity $I_\mathrm{i}=I_\mathrm{t}+I_\mathrm{r}$.
Normally, the light will be almost completely reflected at the first mirror, and only a fraction
finds its way into the cavity (off resonance $I_\mathrm{r}\approx I_\mathrm{i}$).
If however the resonance condition $Mc=2\nu L$ is fulfilled for any positive integer $M$
(longitudinal mode number), the circulating light field builds up inside the cavity over successive
cycles.
Eventually it completely destructively interferes with the reflected light field and the
transmitted intensity equals the incident intensity (on resonance
$I_\mathrm{t}\approx I_\mathrm{i}$).
\par
The frequency difference between two neighboring resonances at $M$ and $M+1$ is called free spectral range (FSR)
and given by $\delta\nu_\mathrm{FSR}=c/(2L)$.
The exact shape of the transmitted and reflected amplitudes and phases can be derived as a
steady-state problem and is included in many standard textbooks~\cite{Siegman.1986}.
For the purposes of this tutorial it is sufficient to discuss the prominent
features as plotted in \fref{fig:cavities}(b) qualitatively:
The transmitted and reflected intensity signals are periodic and produce an identical
Lorentzian line profile for each longitudinal mode in distances $\delta\nu_\mathrm{FSR}$.
Each peak has a certain FWHM $\delta\nu_\mathrm{FWHM}$ which is
approximately (for $R>0.5$) related to the intensity reflectivity $R$ and length $L$ via the
finesse $\mathcal{F}$ like
\begin{eqnarray}
	\mathcal{F}=\frac{\delta\nu_\mathrm{FSR}}{\delta\nu_\mathrm{FWHM}}\approx\frac{\pi\sqrt{R}}{1-R}.
\end{eqnarray}
Consequently, a smaller FWHM is usually attributed to a high reflectivity $R$, when the length
$L$ is governed by the need for a specific FSR.
The phase shift $\Delta\varphi_\mathrm{r}$ of the reflected light field compared to the incident
light field shows a strong dispersive line shape around the resonances.
The finesse $\mathcal{F}$ can be understood as a quality factor of the optical resonance and is
proportional to the mean number of reflections before a photon leaves the cavity.
This is related to the fact, that the cavity field has a certain build-up time before the
transmission reaches the plotted steady state.
Often high finesses and therefore long build-up times are required to achieve extremely narrow
features.
As this would undermine the idea of a fast, high bandwidth feedback response most laser locking
systems are build using the reflected field~\cite{Taubman.2013}. Other shapes of cavities with more
than two mirrors including the circulating bow-tie setup are used in some situations:
The bow-tie configuration for example has a clear geometric separation between
the incident and reflected light field.
However, the presented fundamental theory does not change with the geometry.
\par
An optical cavity made from spherical mirrors also acts as a filter regarding the transversal
mode profile of the incident laser field.
The overall coupling which manifests itself in the peak intensity transmission or the contrast of
the reflected signal does therefore also depend on a transversal mode matching, e.g.\ to a Gaussian
mode profile.
For a high overall coupling to a cavity the cavity must match the longitudinal modes, longitudinal
impedance, and transversal modes~\cite{Kogelnik.1966}. These must be considered during the design
phase as certain geometries and mirror curvatures might demand impracticably large or convergent
beam shapes.
The open-source software \textsc{Rezonator2}~\cite{Chunosov.2022} can be used for Gaussian beam
type cavity calculations, while the other properties can sometimes be only approximated
by the manufacturers, especially if highly reflective thin-film coatings with $R\gg0.99$ are
applied.
\par
In summary, the advantage of cavities as a reference for laser locking lie in their
simplicity, the principle availability at all wavelengths, and their periodicity.
This means that a laser can be locked essentially at any desired frequency (maybe with the
help of offset or sideband locking as discussed below).
They can also have extremely high finesses of $10^6$ and beyond
which makes them specifically suitable for linewidth narrowing~\cite{Izumi.2014}.
The drawback to all these methods is, that they are still based on a relative reference that
can drift over time and needs to be calibrated against an absolute reference.
By using ultralow-expansion (ULE) spacer materials with low coefficients of thermal expansion (CTEs)
inside a \si{\milli\kelvin}-stabilized, vibration isolated vacuum environment it is possible
to reach a limit of material aging and thermal noise of the mirror
substrates~\cite{Numata.2004, Alnis.2008, Millo.2009}.
Possible solutions include the use of different materials with low thermal noise at cryogenic
temperatures~\cite{Matei.2017}.
For practical applications it can also be sufficient to maximize some observed signal
without ever fully calibrating the absolute frequency of the cavity modes.
\subsubsection{Side of fringe lock}
A typical first idea is to \enquote{simply} lock the laser to the peak of an atomic transmission or
a cavity resonance with a setup as shown in \fref{fig:cavities}(a).
While the reflected signal in \fref{fig:cavities}(b) shows a clear minimum to the human
eye, such a signal does not contain an information about the sign of a frequency deviation,
i.e.\ the signal intensity just increases in either direction making it impossible to precisely
counteract.
It is however possible to use such a signal when locking to the side of the resonance, also
called fringe~\cite{Barger.1973, MacAdam.1992}.
The reference is given by a stable voltage $U_\mathrm{set}$ which defines a specific frequency
compared to the photodetector voltage $U_\mathrm{r}(\nu)$ in reflection~\cite{Hamilton.1989}.
This has the immediate advantages of great simplicity and even versatility, as a
lock point can be chosen at an arbitrary frequency along the flank.
\par
Nevertheless, such laser locks are rarely seen in recent publications, since it is the only
presented technique in this tutorial, which directly transfers amplitude noise of the laser
into frequency noise in first order.
The capture range is limited to half the flank of the chosen resonance and a narrower,
more frequency sensitive FWHM increases this issue even more.
Additionally, it is often simply the goal to be on top of the fringe --- especially with
atomic resonances --- rendering a side of fringe lock largely impractical.
\par
Creating such a top of fringe lock error signal can be seen as the underlying challenge to
be solved by all other presented locking schemes.
\subsubsection{Dither lock}
Dithering as a general concept just means that some parameter --- in our case the
laser frequency --- is modulated with some low frequency $f_\mathrm{dither}$.
Note that the name \enquote{dither locking} has no universally agreed usage throughout academic
literature and might refer to a wide range of frequency and phase modulation locking techniques,
sometimes limited to the ones involving a direct modulation of laser current.
For this tutorial by dither locking we mean a rather low $f_\mathrm{dither}$ which is on the order
of the spectral width at high modulation index, such as described in
references~\cite{Hamilton.1989, Taubman.2013} and called wavelength modulation (WM) in
reference~\cite{Supplee.1994}.
This is typically achieved via direct laser current modulation but can also be done
using AOMs.
In this configuration the transmission signal of a (low finesse) cavity can be used as an input,
meaning that the setup looks like the Pound-Drever-Hall (PDH) setup at much lower
frequency (\fref{fig:PDH_HC}(a)).
A lock-in amplifier is then used to detect the change of phase between the modulation signal and
the observed oscillation, which then serves as the error signal for the controller.
\par
The advantage over PDH lies in its lower frequency operation that can be achieved
with much simpler lock-in amplifiers.
The low modulation frequency simultaneously sets a bandwidth limit, which might even be worsened
by the low-pass filtering of a cavity in transmission.
Historically it was chosen in particular since it can be easily modulated through the current input
of any laser, which also leads to an undesired high modulation index of the laser frequency
throughout the rest of the experiment~\cite{Rowley.1963, Salomon.1988, Taubman.2000, Talvitie.1998,
Oxley.2010, Liu.2012}.
\subsubsection{Pound-Drever-Hall (PDH) type lock}
\begin{figure}
	\centering
	\includegraphics{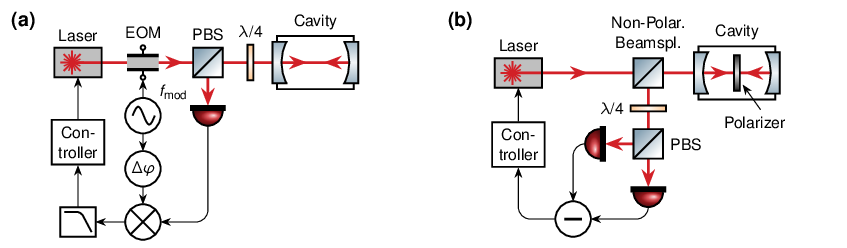}
	\caption{
		(a) The standard configuration for a PDH type laser lock, where the phase modulated
		and reflected beam is detected and mixed (multiplier symbol) with the phase-shifted LO
		(indicated by the sine-wave and $\Delta\phi$ symbol).
		After low-pass filtering the error signal is obtained with is fed into the controller.
		(b) The Hänsch-Couillaud type lock is based on a polarization birefringence of a cavity
		induced by some internal polarizer.
		This is observed by means of the difference signal between two photodetectors.
	}
	\label{fig:PDH_HC}
\end{figure}
The PDH scheme was originally de\-mon\-stra\-ted with masers by Pound in
1946~\cite{Pound.1946} and later applied to lasers in 1983 by
Drever~et~al.~\cite{Drever.1983}.
It appears to be an extension of the dither lock to higher RF modulation
at much lower modulation index at a first glance, but works due to a fundamentally different
effect~\cite{Black.2001}.
\par
A PDH setup as shown in \fref{fig:PDH_HC}(a) consists of a laser, which is phase
modulated at $f_\mathrm{mod}$, reflected by an optical cavity, and observed on
a photodetector with a higher bandwidth than $f_\mathrm{mod}$.
The idea is to recover the phase information of the reflected signal
(\fref{fig:cavities}(b)), which is only possible by an interferometric technique at optical
frequencies.
The modulation adds sidebands at $\nu\pm f_\mathrm{mod}$ that also get reflected at the cavity.
Since they lie away from the resonance, they are essentially completely reflected with no
significant phase shift and therefore serve as a phase reference of the heterodyne
interference signal.
The carrier frequency $\nu$ gets only partially reflected with a strong
phase shift imprinted for small deviations from the resonance frequency.
It is this phase difference in interference with the virtually unchanged sidebands
that leads to an amplitude modulation on the photodetector at
$f_\mathrm{mod}$, which can be recovered electronically by mixing (i.e.\ multiplying) with
a reference $f_\mathrm{mod}$ signal.
After a low-pass filter ($<f_\mathrm{mod}$) a dispersive error signal is created and
subsequently processed by the following controller.
\par
One immediate advantage of the PDH scheme is its inherent balance or \enquote{nulling} around
the cavity resonance, as both sidebands  $\nu\pm f_\mathrm{mod}$ contribute equally to the signal.
This ideally results in a signal with no offset, which is also not susceptible to intensity
fluctuations of the laser.
Furthermore, the detection at $f_\mathrm{mod}$ occurs in a regime much higher than common
low-frequency technical noise sources~\cite{Taubman.2013}.
Only a small modulation index is necessary for a strong interference as only a small portion
of the carrier gets reflected~\cite{Black.2001}.
The capture range of the method is set by $f_\mathrm{mod}$ in either direction from the cavity
resonance, at which point the sidebands themselves will be on resonance and introduce another zero
crossing.
\par
The primary disadvantage of the PDH scheme lies in its mandatory RF electronics
and modulation.
PDH detection additionally suffers from any residual amplitude modulation (RAM), that will
produce an offset of the resulting error signal.
Great lengths are taken to minimize RAM, which is
commonly introduced by EOMs, or make it at least a stable contribution over
time~\cite{Zhang.2014, Fang.2017}.
Nevertheless, PDH is probably the most widely adopted technique presented in this tutorial,
both historically and today.
\subsubsection{Hänsch-Couillaud type lock}
Slightly before Drever and Hall published the PDH scheme, Hänsch and Couillaud
introduced another, polarization based frequency discrimination technique in 1980~\cite{Hansch.1980}.
Interestingly, their work is also motivated by the 1946 publication of Pound~\cite{Pound.1946}
while being aware of the work by Drever and Hall from private communication~\cite{Hansch.1980}.
\par
It is therefore not surprising that, similar to the PDH scheme, the Hänsch-Couillaud setup
consists of a cavity at which the laser beam is reflected while the reflected signal is observed
(\fref{fig:PDH_HC}(b)).
In difference however, the cavity must contain a linear polarizer or some optical component which
introduces different properties depending on polarization such as a Brewster-cut or birefringent
crystal.
This effectively makes the cavity as whole birefringent, leading to a phase shift of the reflected
light orthogonal to the axis of the internal polarization optic.
On resonance, both components stay in phase, leading to an effective linear polarization of the
reflected light, while off-resonant light will lead to an elliptic polarization.
Importantly, the handedness of the ellipticity depends on the sign of the detuning between laser
and cavity resonance~\cite{Hansch.1980}.
Since this ellipticity needs to be detected, the reflected signal can not be separated through the
quarter wave plate and polarizing beam splitter (PBS) combination.
\Fref{fig:PDH_HC}(b) proposes a separation using a non polarizing beam splitter
(thus losing at least \SI{75}{\percent} intensity), while
recent applications are mostly found using ring cavities that exhibit a geometric splitting
between incident and reflected beam \cite{Bateman.2010, Vainio.2011, Hannig.2018}.
The detection is performed as balanced detection between two photodetector voltages behind a
quarter wave plate and PBS combination.
This difference does not depend on the azimuth angle of the ellipticity, but only on its
magnitude and handedness~\cite{Hansch.1980}.
It can therefore be directly used as the error signal for the controller.
\par
The major advantage of the Hänsch-Couillaud type lock over PDH is the absence of any RF
components and sideband modulation.
Even though any sidebands would be strongly suppressed in transmission, this makes it the method
of choice in applications where no sidebands can be present at all, e.g.\ during second harmonic generation (SHG)
in certain experiments~\cite{Hannig.2018}.
Due to the balanced operation between two photodetectors, common mode intensity noise is suppressed.
\par
These two photodetectors on the other hand must stay precisely matched to prevent offsets of
the error signal \cite{Hamilton.1989}.
The \enquote{nulling} suffers from bad transversal cavity coupling.
Every component inside the cavity increases losses and reduces its finesse~\cite{Taubman.2013}.
Compared to PDH, the detection occurs at a frequency of zero, which typically features the
largest inherent technical noise.
Most importantly however, it lacks the wide capture range of PDH, rendering locks to narrow
cavity lines particularly challenging.
\subsection{Locking to an atomic or molecular reference}
While the previous section on cavities inevitably required the discussion of the stability
and stabilization of such devices themselves, atomic or molecular transitions are seen as the
ultimate source of frequency stability and absolute precision.
Locking a laser to an atomic resonance instead of a cavity works not fundamentally different either.
Much rather, many of the previously stated ideas and schemes can be directly applied (and get
different names nevertheless).
This however comes at the cost of versatility: Resonances can only be found at specific frequencies
and come in fixed strengths and linewidths.
Moreover, atomic resonances are not as stable as they seem to be at the first glance, being
influenced by external electric and magnetic fields, other atoms in their vicinity, as well as their
own movement~\cite{Taubman.2013}.
Consequently, laser locking techniques using atomic standards can be roughly divided into two
categories: First, there are compact vapor cell based devices used to produce light which
is to be interfaced with another sample of the same atomic species.
In this configuration the described linewidth and versatility limits are actually advantageous,
since they are intrinsically suitable for the application.
The second category includes typically room-filling experiments which opt to set frequency
standards based on optical transitions in state-of-the-art optical atomic
clocks~\cite{Ludlow.2015}.
For the purposes of this tutorial, we primarily have the first category in mind, even though
the fundamental techniques are not strictly different but rather more elaborate in the second case.
All schemes discussed in the following section broadly fall under the umbrella of
laser spectroscopy techniques which are detailed in textbook
literature such as the ones by Demtröder~\cite{Demtroder.2008, Demtroder.2015}.
\subsubsection{Saturated absorption spectroscopy (SAS) dither lock}
Of the techniques presented in \sref{sec:cavitylock}, both side-of-fringe and dither locking
can be directly applied to an atomic resonance observed in transmission through a reference vapor
cell~\cite{Sakai.1992}.
This configuration however suffers from inherent Doppler broadening, especially if additional
heating beyond room-temperature is required to achieve sufficient vapor densities.
Saturated absorption spectroscopy (SAS) refers to a setup, where the incident laser beam is split
into a probe and a pump beam, which then cross the vapor cell in a counter-propagating
fashion~\cite{Lee.1967, Haroche.1972, Preston.1996}.
\Fref{fig:PS_DAVLL} shows two methods derived from SAS with additional polarization optics,
whereas the most basic form requires no additional polarizers and just one photodetector.
Doing so creates so-called sub-Doppler Lamb dips in the transmission spectrum at the frequency
of the atomic resonances.
Their specific shape and linewidth is subject to beam-diameters, temperatures, angles, and external
fields \cite{Haroche.1972, Schmidt.1994}.
\par
Whilst SAS is most commonly just understood as a spectroscopy method, it is
occasionally referred to as a locking technique on its own in scientific publications.
If done so, this mostly refers to a dither lock by means of laser current modulation
applied around a Lamb dip achieved through SAS \cite{Ohta.2006, Wang.2015b, Genov.2017}.
Staying in line with the naming conventions, the name wavelength modulation spectroscopy (WMS)
is also used to differentiate it from frequency modulation spectroscopy (FMS) described
below~\cite{Fang.2017}.
\par
SAS and all derived schemes presented below can often be applied to pairs of pump and probe
lasers at different frequencies in very similar configurations. This is useful to lock lasers
which are not directly coupled to a ground state.
\subsubsection{Polarization spectroscopy lock}
\begin{figure}
	\centering
	\includegraphics{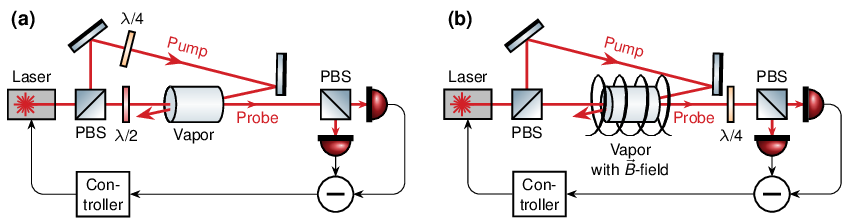}
	\caption{
		Two modulation free SAS type locking schemes using counter-propagating pump and probe
		beams to induce an anisotropy in the atomic medium.
		(a) The polarization spectroscopy lock induces a birefringence in the atomic media using a
		circular polarized pump.
		(b) The DAVLL type lock uses a magnetic field to induce a dichroism due to Zeeman shifts.
	}
	\label{fig:PS_DAVLL}
\end{figure}
Of the methods derived from SAS, polarization spectroscopy was introduced by Wieman and Hänsch
in 1976~\cite{Wieman.1976}.
It can be understood as a technique similar to the Hänsch-Couillaud scheme using an induced
polarization birefringence inside a vapor sample instead of a cavity.
The setup as shown in \fref{fig:PS_DAVLL}(a) consists of an SAS configuration which is modified
using a circular polarized pump beam combined with polarization dependent separation and
differential detection of the probe beam in transmission.
A much stronger pump beam introduces a birefringence inside the atomic medium due to an
anisotropic population of $m$ states from pumping with a circular polarization~\cite{Pearman.2002}.
This leads to an ellipticity of the probe beam behind the vapor sample which changes sign at the
resonance and increases angle and magnitude roughly proportional with
detuning~\cite{Harris.2006}.
\par
The difference signal features a steep dispersive shape around resonance that can be tuned by
changing beam powers or diameters and then directly be used as an error
signal~\cite{Pearman.2002, Harris.2006, Do.2008}.
Many advantages and drawbacks are shared with the Hänsch-Couillaud scheme:
Common mode intensity noise on the laser beam is rejected between the two photodiodes and
a minimal absorption in a thin or low-pressure sample which is nevertheless accompanied by a strong
dispersion can lead to a sufficient SNR~\cite{Yoshikawa.2003, Groswasser.2009}.
No modulation of the laser beam is required, and a larger capture range compared to
SAS dither spectroscopy is usually possible~\cite{Pearman.2002}, thus featuring
a remarkable performance given its simplicity.
A side effect of the underlying anisotropic pumping is that cycling transitions (where the
excited state can not decay into a different state) are strongly enhanced over non-cycling
transitions and cross-over resonances~\cite{Torrance.2016}.
\par
The technique also suffers from all drawbacks it shares with the other SAS based
and non-modulated locking variants:
The capture range is inherently limited by the linewidth of the interrogated
transition.
Moreover, the locking point is very sensitive to external fields~\cite{Groswasser.2009}
and the actual atomic resonance is usually not located at the center of the dispersive slope
due to a strong Doppler background~\cite{Pearman.2002}.
Some of these issues can be circumvented by sufficient shielding or a slightly modified detection
process~\cite{Ratnapala.2004, Tiwari.2006, Kunz.2013} such that polarization spectroscopy locks
are successfully applied in various recent publications~\cite{Lee.2014, Torrance.2016, Wang.2022}.
\subsubsection{Dichroic atomic vapor laser lock (DAVLL)}
Another approach to introduce an anisotropy into a vapor sample relies on Zeeman shifts through
externally applied magnetic fields.
Different schemes including modulated and static magnetic fields
have been proposed~\cite{Weis.1988, Cheron.1994, Yashchuk.2000}, of which the
dichroic atomic vapor laser lock (DAVLL) remains the seemingly
most widespread variant since its publication in 1998~\cite{Corwin.1998} and extension to SAS
around 2003~\cite{Wasik.2002, Petelski.2003}.
\par
\Fref{fig:PS_DAVLL}(b) depicts the principle setup, which, being another SAS based method,
shows noticeable similarities to the polarization spectroscopy lock.
Both indeed rely on a balanced detection, but DAVLL is typically understood to be probing
the dichroic $\sigma^\pm$ components using a quarter wave plate and PBS combination instead of
the  ellipticity birefringent measurement described above.
Since the Zeeman splitting lifts the degeneracy of different $m$ states, the method effectively
probes the difference in transmission between all $\Delta m=-1$ and $\Delta m=+1$ transitions.
On resonance, both are identical, while any detuning will lead to a stronger contribution in either
direction.
The resulting difference signal is again dispersive and might be optimized using the magnetic field
strength~\cite{MillettSikking.2007}.
\par
The DAVLL scheme again shares many of the advantages and drawbacks with polarization spectroscopy
locking, including its simplicity, the lack of modulation at cost of high technical noise,
limited capture range, Doppler background, and sensitivity to external fields.
The stability and thermal influence of any magnetic field coil can be
challenging~\cite{Reeves.2006, McCarron.2007}, but some effort went into optimizing,
embedding, or miniaturizing DAVLL based locking systems in recent
years~\cite{Harris.2008, Lee.2011, Su.2014, Pustelny.2016}.
\subsubsection{Frequency modulation spectroscopy (FMS) lock}
\begin{figure}
	\centering
	\includegraphics{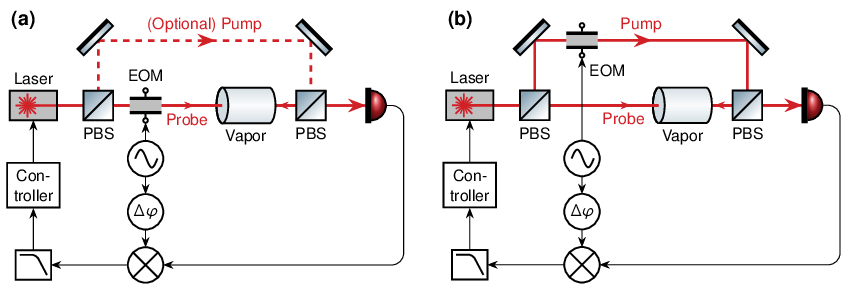}
	\caption{
		Schematics of two common methods to create an error signal using RF modulation and an
		atomic vapor.
		(a) In the FMS scheme, the probe beam gets directly modulated, such that the
		dispersion of the sample can be extracted by a frequency mixer.
		(b) In the MTS scheme, the pump gets modulated leading to a transfer of this modulation
		to the probe beam.
	}
	\label{fig:FMS_MTS}
\end{figure}
The idea to lift the error signal generation far away from the low frequency floor of
technical noise was not only explored in the context of PDH with optical cavities, but
also as FMS of atomic media throughout the
1980s~\cite{Bjorklund.1980, Schenzle.1982, Bjorklund.1983} (based on initial work from the
1960s~\cite{Harris.1965}).
Similar to the PDH technique, the probe beam gets phase modulated to generate sidebands
$\nu\pm f_\mathrm{mod}$ which heterodynely interfere and form a modulated signal on a
photodetector (shown in \fref{fig:FMS_MTS}(a) including an optional pump beam to create
sub-Doppler features).
This contains information about both the absorption and dispersion, the latter of which is
demodulated using a mixer to create a literally \enquote{dispersive} error signal.
By tuning the modulation frequency and modulation index with respect to the natural linewidth of the
transition, the shape of this signal can be tweaked for specific locking (or measurement)
purposes~\cite{Bjorklund.1983, Supplee.1994, Cornish.2000}.
Inheriting the drawbacks and advantages already described for PDH type lock, FMS suffers
from additional asymmetries and offsets arising from nearby transitions~\cite{So.2019}, which
do not exist in a perfectly periodic optical cavity.
Being an RF modulation technique it nevertheless lowers the overall noise floor and
reduces offset drifts~\cite{Zi.2017} and can therefore be found in various
recent experiments and atomic species~\cite{Bell.2007, Norcia.2016, Jrgensen.2016, Zi.2017,
Spindeldreier.27.01.201801.02.2018}.
\subsubsection{Modulation transfer spectroscopy (MTS) lock}
Modulation transfer spectroscopy (MTS) is closely related to FMS to the point where it can be
nontrivial to identify either scheme in specific cases, while even mixed
variants have been described~\cite{Zi.2017}.
The spectroscopic technique was first described in the early 1980s~\cite{Snyder.1980, Raj.1980,
Shirley.1982, Ducloy.1982} and applied to locking not later than 1995~\cite{Eickhoff.1995}
(reference~\cite{Sun.2016} briefly outlines the history).
\Fref{fig:FMS_MTS} shows an MTS locking setup, which is remarkably similar to FMS locking,
yet different in that the pump beam is modulated in lieu of the probe beam.
The eponymous \enquote{transfer} of modulation from the pump beam to the probe beam ultimately leads
to a modulation on the photodetector that, again, can be demodulated creating a
dispersive error signal.
Due to the transfer of modulation, the underlying physical mechanism is a four-wave mixing process
in which two probe photons at $\nu$ and one pump photon at $\nu\pm f_\mathrm{mod}$ form
a signal photon at $\nu\mp f_\mathrm{mod}$.
These photons then interfere with the probe photons and induce a modulation at $f_\mathrm{mod}$
on the photodetector.
\par
In difference to FMS the resulting error signal has a flat, \enquote{nulled} background
where the resonance peaks coincide with the zero crossings~\cite{Ducloy.1982}.
This
is not affected by laser intensity, laser polarization, and temperature
in first order~\cite{Sun.2016}.
MTS, in contrast to FMS, is usually only applied with modulation frequencies smaller
than the natural linewidth, since this produces the steepest, most sensitive slope in the error
signal~\cite{Ducloy.1982, Bertinetto.2001, McCarron.2008, Zhang.2003} but limits the possible
feedback bandwidth.
Tuning this steepness might be a desired property which can be influenced by geometric
beam parameters, modulation index, and polarization~\cite{McCarron.2008, Noh.2011}.
Similar to polarization spectroscopy, only cycling transitions create a significant MTS signals
while crossover resonances are suppressed~\cite{Noh.2011}.
RAM can be introduced by MTS itself and adversely affects its locking performance if
present~\cite{Jaatinen.2009}.
\subsection{Locking to a reference laser}
It is a natural question to ask whether it is possible to use an already stabilized laser
as an optical oscillation \enquote{reference} for another laser, called \enquote{target}.
This is particularly plausible in applications which require multiple
lasers with possibly $<\SI{10}{\giga\hertz}$ frequency difference that interact with the same
atomic species.
Doppler cooling of atoms using a repumping laser typically fulfills these criteria.
There are even constellations like coherent population trapping, where only the
difference frequency between two lasers must be stabilized while the exact absolute frequency
of both components is less critical~\cite{Shah.2010, Shalagin.2017}.
\subsubsection{Transfer cavity locks}
\begin{figure}
	\centering
	\includegraphics{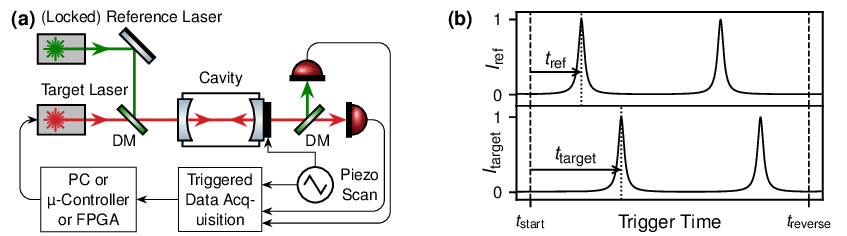}
	\caption{
		(a) Schematic drawing of a scanning transfer cavity stabilizing a target laser to a
		reference laser by observing the transmission of both triggered by the piezoelectric modulation
		of the cavity length.
		(b) The signals captured by the triggered acquisition show the periodic transmission
		peaks of the cavity, which are reflected about each typically triangular signal ramp starting
		at $t_\mathrm{start}$ and ending at $t_\mathrm{reverse}$.
		The feedback loop acts on the time difference between $t_\mathrm{target}$ and $t_\mathrm{ref}$.
	}
	\label{fig:ScanTransfer}
\end{figure}
Based on the discussions of cavity based locking in \sref{sec:cavitylock}, a \enquote{transfer}
locking scheme can be directly derived:
Couple the reference laser to a cavity that can be length-tuned using a piezoelectric actuator
and employ any of the aforementioned techniques to lock the cavity length to a laser resonance.
This requires slightly different feedback loops, since piezoelectric actuators fundamentally lack
any tuning faster than their mechanical properties in the \si{\kilo\hertz} range.
The locked cavity can then be used as a stabilized frequency reference for the target laser, even
if e.g. the temperature changes or slow vibrations are applied~\cite{Riedle.1994, Uetake.2009,
BohlouliZanjani.2006, Biesheuvel.2013}.
A drawback of this method lies in the chaining of individual locks,
each of which potentially introduces noise into the system.
It however can be built much more compact than a ULE cavity inside a vacuum chamber, and has
the opportunity to be virtually infinitely expanded using just one reference laser.
If the cavity is kept under atmospheric pressure, differential refractive index changes between
the two wavelengths must be corrected~\cite{Subhankar.2019}.
The overlapping and separation of reference and target beam can be achieved based on polarization
or wavelength (using e.g. a grating or dichroic mirror).
A specific transfer cavity system as used in our laboratory is discussed in \sref{sec:examplelock}.
\par
Another related and in some sense even more simplified approach relies on continuously modulating
the cavity length (called scanning) using the piezoelectric actuator on a \si{\hertz} to
\si{\kilo\hertz} timescale~\cite{Lindsay.1991, Zhao.1998, Rossi.2002}.
In this scanning transfer cavity scheme as depicted in \fref{fig:ScanTransfer}(a) it becomes
possible to observe a low finesse transfer cavity in transmission and detect both target and
reference beam triggered against the scanning ramp generator.
The captured signals (\fref{fig:ScanTransfer}(b)) will exhibit a certain time difference
$t_\mathrm{target}-t_\mathrm{ref}$ that encodes the frequency difference between both lasers.
By keeping this temporal distance constant, again correcting for refractive index changes,
the frequency of the target laser can be stabilized to any desired cavity resonance.
The scanning transfer technique is strictly bandwidth limited by the mechanical action of the
piezoelectric actuator, even if it is driven at mechanical resonance~\cite{Tonyushkin.2007}.
This softens the requirements towards the control electronics and allows the implementation
of affordable PC or microcontroller based locking system, which evaluate the signals captured
by the photodiodes~\cite{Burke.2005, SeymourSmith.2010, Subhankar.2019}.
It also avoids locking the cavity length as an intermediate step.
Overall, based on the low bandwidth and long delays, the scanning transfer cavity technique
is not employed where linewidth narrowing is desired, but only, when an absolute frequency stability
with the free-running linewidth of the laser is already sufficient.
\subsubsection{Optical frequency lock loops (OFLL)}
\begin{figure}
	\centering
	\includegraphics{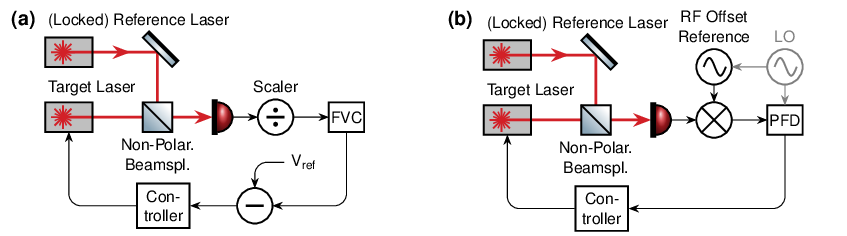}
	\caption{
		Schematics of two locking schemes derived from the beat note between two lasers on a
		photodetector.
		(a) In an OFLL this beat oscillation is then scaled (divided by an integer) and converted
		to a voltage by an FVC. The error signal is obtained by subtracting a certain reference
		voltage $V_\mathrm{ref}$.
		(b) The OPLL mixes the beat oscillation with an RF reference oscillation, that
		governs the frequency offset. By comparing frequency and phase in a PFD that is clocked
		by the same LO, the error signal is generated.
	}
	\label{fig:OFLL_OPLL}
\end{figure}
Optical frequency lock loop (OFLL) and the closely related optical phase lock loop (OPLL),
both of which are sometimes referred to as \enquote{offset locking} or \enquote{beat note locking},
stabilize a target laser at an RF scale frequency offset to a reference laser.
Historically, the OPLL as derived from electronic phase lock loops (PLL) was explored much
earlier in the 1960s~\cite{Enloe.1965, Preveddelli.1995} compared to the first experiments with
OFLLs in the 1990s~\cite{Stace.1998, Schunemann.1999}.
In both cases, the two lasers are brought to interference on a photodetector, creating a beat note
oscillation of which only the low frequency difference
$|\nu_\mathrm{reference}-\nu_\mathrm{target}|$ will be directly electronically captured
(requiring a fast enough photodetector).
This frequency component is then fed through a frequency to voltage converter (FVC) and compared
to a reference voltage $V_\mathrm{ref}$ (\fref{fig:OFLL_OPLL}(a)).
To increase the accessible RF frequency range, a (variable) frequency scaler, that reduces the
frequency by an integer, is routinely included in such applications.
An OFLL error signal often features a wide capture range that can be employed to lock lasers
with high phase noise, low bandwidth feedback loops, and low
SNR~\cite{McFerran.2018, Reynolds.2019}.
This is advantageous if the lock can track a fast sweep of the reference voltage that leads
to a fast and unambiguous sweep of the target laser frequency~\cite{Strau.2007, Wang.2022}.
Various other architectures based on delay lines or RF filters in combination with an RF
oscillator, which replace the scaler and FVC discriminator, have been proposed in
the literature to overcome their limited frequency range and enhance the overall OFLL
performance~\cite{Ritt.2004, Schilt.2008, Uehara.2014, Lipka.2017, Li.2022}.
\subsubsection{Optical phase lock loops (OPLL)}
Certain types of experiments, e.g. atom interferometry, require phase coherence between multiple
lasers.
If the phase between two lasers stays perfectly synchronized --- even if their frequencies are at an
offset --- their frequencies will also stay in perfect distance.
The principle OPLL setup in \fref{fig:OFLL_OPLL}(b) illustrates the underlying idea to compare
the beat note frequency with a known stable RF reference oscillator.
Compared to an OFLL, such an OPLL demands a much higher feedback bandwidth to keep the
phase slip within the narrow capture range of $\SI[parse-numbers=false]{\pi/2}{\radian}$ at all
times (beyond which the mixer becomes periodic)~\cite{Taubman.2013, Lipka.2017}.
Phase frequency detectors (PFDs) improve the capture range by comparing both phase and frequency and
pulling the laser towards the OPLL capture range, if the phase difference gets too large.
Choosing an exclusive phase comparison over a PFD, i.e. by low-pass filtering directly behind
the mixer, is possible, but only feasible for laser systems with intrinsically narrow linewidths and
high feedback loop bandwidth~\cite{Preveddelli.1995}.
In the context of this tutorial, the OPLL typically requires the highest feedback bandwidth of
all systems.
Consequently, a wide variety of publications aiming to soften the requirements, reducing
cost~\cite{Hockel.2009, Ivanov.2011, Kawalec.2013}, or shifting the work to an all-digital
PFD exist~\cite{Cacciapuoti.2005, Wang.2011, Xu.2012}.
\subsubsection{Optical frequency combs (OFC)}
\begin{figure}
	\centering
	\includegraphics{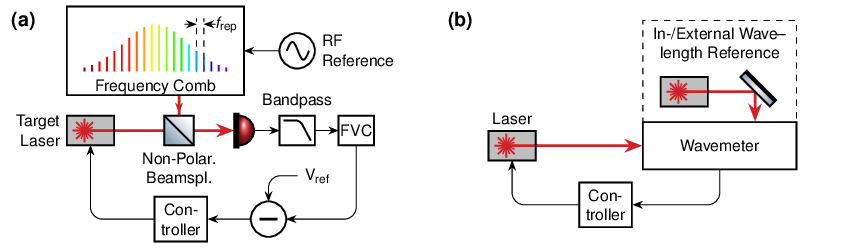}
	\caption{
		(a) An OFC generates a series of well defined optical frequencies which are fully determined
		based on two electrical RF measurements. Each of those frequencies can subsequently be used
		in a OPLL or OFLL like drawn here.
		(b) A wavemeter can directly measure the exact frequency of the laser based on one or more
		internal interferometers. For high absolute precisions this requires regular calibration
		based on a known stable light source.
		}
	\label{fig:OFC_Wavemeter}
\end{figure}
An optical frequency comb (OFC), if used as a reference laser system, can solve the absolute
frequency issue.
Its optical oscillation modes are completely determined by two RF oscillators,
which can in turn be locked to an atomic clock.
Yet, it would be an understatement to just call it another laser system, considering its deeply
connected history with frequency metrology, laser locking, and pulsed lasers as
outlined by Hänsch in his Nobel Prize lecture (received for the development of OFCs and
contributions to precision laser spectroscopy)~\cite{Hansch.2006}.
Femtosecond OFCs rely on the fundamental --- yet technically not easily exploited ---
Fourier relationship between temporal and frequency components of a mode-locked pulsed laser.
Mode-locking inside a non-linear cavity leads to a tight temporal relationship between neighboring
frequency modes, enabling to the emission of a train of short pulses~\cite{Cundiff.2003,
Diddams.2010, Fortier.2019}.
If the emitted spectrum is sufficiently broad and the dispersion adequately controlled, the pulses
with repetition rate $f_\mathrm{rep}$ necessarily yield a harmonic frequency comb with a
longitudinal mode spacing $f_\mathrm{rep}$, that can be controlled through the cavity length.
The frequency of the $N$-th longitudinal mode is governed by the \enquote{comb equation}
\begin{eqnarray}
	\nu_N=Nf_\mathrm{rep}+f_0,
\end{eqnarray}
where $f_0$ is an additional, randomly drifting offset frequency generated by a phase shift between
successive pulses.
Measuring and fixing $f_\mathrm{rep}$ is straight forward by cavity length tuning,
while accessing $f_0$ is the innovation that enables
building a truly self-referencing OFC.
If the spectral bandwidth of the comb spans more than an octave --- that means two frequency
components exist where some $\nu_{2N}=2Nf_\mathrm{rep}+f_0$ have at least twice the frequency
of some other $\nu_N$ --- this lower frequency can be doubled and brought to interference with
$\nu_{2N}$ to measure the beat note $2\nu_N-\nu_{2N}=f_0$.
Now, an intra-cavity dispersive optical element changes the group delay and stabilizes $f_0$,
thus fixing all $\nu_N$ precisely based on two RF frequency measurements.
\par
Locking a laser to an OFC is usually performed using an OFLL or an OPLL, which
additionally selects a single frequency component of the comb (often the closest) by means of
an electrical band-pass filter~\cite{Jost.2002}.
The ambiguity or discontinuity when tuning the laser frequency across multiple comb frequencies can
be overcome, such that fast and wide (many \si{\nano\meter}) frequency sweeps locked to an OFC
have been demonstrated~\cite{Schibli.2005, Park.2006, Inaba.2006, Prehn.2017}.
Interest in frequency locking with OFCs is particularly strong in precision metrology, due to
their absolute frequency standard and possible narrow-line phase
locks~\cite{Akamatsu.2012, Argence.2015}.
For applications with cooled atoms or hot atomic vapors, an OFC can form the shared single
source of stability for many laser systems at a range of wavelengths~\cite{Yasuda.2010,
Fordell.2014, Gunton.2015, McFerran.2018}.
This can offset the high upfront cost that is usually associated with an OFC, yet depending
on the spectral range multiple OFCs might be necessary, if they are available at all.
\subsection{Locking to an interferometer or wavemeter}
Optical wavemeters are routinely used with single frequency lasers to set and monitor wavelengths.
This is particularly true in combination with any of the presented frequency discrimination
techniques, e.g. to verify that the correct mode of a cavity or an OFC has been selected.
These devices are really interferometers of different kinds, such as
scanning Michelson interferometers~\cite{Kowalski.1977, Fox.1999},
multiple Fabry-Pérot interferometers~\cite{Scholl.2004},
and multiple Fizeau interferometers~\cite{Castell.1985, Gardner.1985, Gray.1986} in
combination with fast acquisition, signal processing, and sufficient calibration.
\par
Since wavemeters with a relative precision of $10^{-8}$ and
better (corresponding to less than \SI{10}{\mega\hertz}
absolute precision in the visible spectrum) became commercially
available~\cite{Verlinde.2020, Konig.2020}, they have been used to stabilize the
frequency of lasers~\cite{Kobtsev.2007}.
Wavemeters offering this precision require frequent calibration and are usually operated in
combination with a known stable reference light source (\fref{fig:OFC_Wavemeter}(b)).
The biggest advantage of wavemeters lies in their virtually infinite capture range, that is
far wider than the emission band of most laser sources, and the possibility to lock at any
frequency within this range without additional components or modulation.
Due to the signal processing involved, wavemeters typically have acquisition rates
of less than \SI{1}{\kilo\hertz}, and therefore offer only limited feedback
bandwidth, essentially ruling out any linewidth narrowing.
As high precision wavemeters also come at increased cost, it is common to use them in
combination with fast optical switches and multiple target lasers
simultaneously~\cite{Couturier.2018, Saakyan.2015}.
This limits the acquisition rate for each individual target laser even more, but does not strongly
impact the locking performance, as only a suppression of frequency drifts is possible anyway.
Research groups have successfully demonstrated the use of such systems for Doppler cooling and
trapping of various atomic species~\cite{Ghadimi.2020, Kim.2021}.
\subsection{Common extensions and other types of frequency locks}
Complex experiments often rely on different laser locks for different
systems, e.g. a wavemeter lock for a series of broad ground state transitions, supplemented
by narrow linewidth OFC locks or even extremely narrow PDH cavity locks.
Locking a single laser using multiple of the presented frequency discrimination schemes is
necessary, e.g. if two seemingly incompatible requirements arise, i.e. locking a noisy laser to a
narrow linewidth cavity but with outstandingly large capture range or stability.
Some techniques like PDH and FMS require an RF modulation, but can only be applied
directly at a resonance.
Sideband locking overcomes this limitation by modulating most of the laser energy into a sideband
at the desired offset and locking to this sideband~\cite{Thorpe.2008}.
Finally, electronic feedback might be avoided completely in favor of purely optical feedback beyond
the internal laser cavity, e.g. from an external cavity (as done in ECDLs), from an atomic
transition~\cite{Cuneo.1994}, or injection locked from another laser or
OFC~\cite{Buczek.1973, Lang.1982, Park.2006}.
\section{An exemplary locking scheme}
\label{sec:examplelock}
Given the wide range of possible applications with hot atomic vapors alone, it is impossible to
give universal advise on designing a laser lock in the style of a generally valid recipe.
In this section we will much rather study one system that was developed and extended in our
laboratories during the past years.
We will focus particularly on the process that has led us to settle with and eventually build
the setup as it stands, while covering some more general aspects that were omitted in the
previous sections.
\subsection{Decision Process}
Our research activities using thermal atomic vapors include vapor-cell manufacturing and testing,
electromagnetically induced transparency (EIT) type schemes involving the Rydberg blockade,
nanosecond four-wave mixing in microscopic vapor-cells, gas sensing applications, and
integration of photonic chips with thermal atomic vapors.
Currently, cesium atoms, rubidium atoms, and nitric-oxide molecules are the primary subject of
research, but other atomic and molecular species have been used or might be used in the future.
In contrast to many experiments involving cooled or trapped atoms that require specifically
designed vacuum chambers, Zeeman slowers, or magnetic field coils, it is feasible to
perform studies involving different atomic species (e.g. Cs and Rb) and switching between those in
\enquote{quick} succession by exchanging vapor-cells.
To facilitate such research, we operate a set of shared lasers, e.g. the alkali D-line
transitions and their successive Rydberg transitions, which are distributed from a central,
rarely entered, temperature stabilized room using polarization maintaining fibers.
These are supplemented by laser systems more specific or tailored to particular setups
located close to their intended application, such as pulsed amplifiers, unusual excitation
paths, or high power beams.
\par
The setup as previously operated was highly heterogeneous, containing numerous self-made ECDLs,
commercial ECDLs, and SHG cavities locked through a mixture of commercial and self-made
PID controllers.
Optical frequency discrimination was performed primarily against atomic standards using
DAVLL and dither locking but also using PDH with a ULE cavity.
Since the students and researchers, who had originally conceived those laser locks, mostly had left
for different laboratories and the electronics increasingly showed aging-related failures,
a replacement was sought for the next generation of laser systems.
Our requirements and informal wish-list for laser locking therefore included:
\begin{itemize}
	\item Direct locking of at least 12 lasers at a wide range of wavelengths between
				\SI{420}{\nano\meter} (Rb $5S_{1/2} \to 6P_{1/2}$)
				and \SI{1529}{\nano\meter} (Rb $5P_{3/2} \to 4D_{5/2}$), including multiple
				lasers at identical or almost identical wavelengths.
	\item Linewidth of the lasers should be on the order of few \SI{100}{\kilo\hertz}
				(few \si{\kilo\hertz} linewidth might be necessary in the future).
	\item Locking is possible with large, arbitrary detuning from atomic transitions (some planned
				excitation schemes require hundreds of \si{\giga\hertz} detuning from atomic resonance,
				others use weak transitions between high NO molecular states).
	\item Ideally, a single solution exists for all laser locks.
	\item Easy and cost-effective extensibility and scalability for additional laser systems.
	\item Electronics are readily available for purchase or reproduction (and can ideally be modified
				for other feedback applications such as intensity stabilization).
	\item Monitoring and remote control possible via our institute's Ethernet network during
				experimentation in neighboring rooms.
	\item Digital electronics are generally favorable since locking settings can be
 				stored and reproduced.
	\item Ideally open-source software gets employed for most tasks.
	\item Possibility to continue using all existing laser systems.
	\item Achieves a homogeneous way to introduce new students to the locking system.
\end{itemize}
Locking lasers at arbitrary frequencies essentially rules out any atomic vapor reference,
which would have been only possible in some of our use-cases.
A wavemeter or OFC offers a reasonable turn-key solution, but turned out to
require multiple of the most-expensive devices commercially available.
It also lacks cost-effective gradual scalability as new wavelengths or a larger number of
lasers might require additional, expensive hardware.
\par
We therefore decided to design a cavity based system, which could additionally serve as a frequency
ruler during scanned operation of lasers.
A ULE cavity with an FSR of roughly \SI{1.5}{\giga\hertz} and a finesse of about
\num{3000} at certain wavelengths for rubidium spectroscopy was already available.
The purchase of additional ULE cavities to be used with a set of lasers overlapped through
dichroic mirrors was considered but impractical due to high upfront cost, complexity, and limited
extensibility, if e.g. multiple laser need to be operated at adjacent frequencies.
\par
Eventually the decision came towards transfer-cavity locking, with a reference laser either
locked to a Rb $5S_{1/2} \to 6P_{3/2}$ hyperfine ground state transition using DAVLL or to
a ULE cavity using PDH.
The former leaves the ULE cavity unoccupied for other applications, but makes transfer to
another laser at the same atomic transition impossible with a dichroic mirror, while the latter
permits linewidth narrowing.
Modulation on the reference laser can be introduced through current modulation, as it does not
serve any purpose beyond stability transfer.
Stabilizing the target lasers might either be done by locking the transfer cavity length using
the same modulation and PDH, if a suitable modulator is available for the target laser, too.
If not, the same cavity can be operated as scanning transfer system.
The standardized optical setup around self-built
cavities (with about \SI{1}{\giga\hertz} FSR and a medium finesse of about \num{1000}) is
clearly and logically separated from the feedback electronics, which can be upgraded or exchanged
if required.
\par
A drawback of this architecture lies in the large number of frequency locks and the
spatial separation of up to \SI{30}{\meter} between
the reference laser and some specific laser systems in other rooms.
While it is possible to locate the transfer cavity on either end, RF signals must be
transferred for some PDH lock and feedback over this distance.
Additional care must be taken to avoid any form of cross-talk between the different locks
in both the optical and electronic domain.
Our colleagues in a neighboring lab took a similar approach based on scanning transfer cavities as described in \cite{Pultinevicius.2023}.
\subsection{Setup}
\label{sec:oursetup}
\begin{figure}
	\centering
	\includegraphics{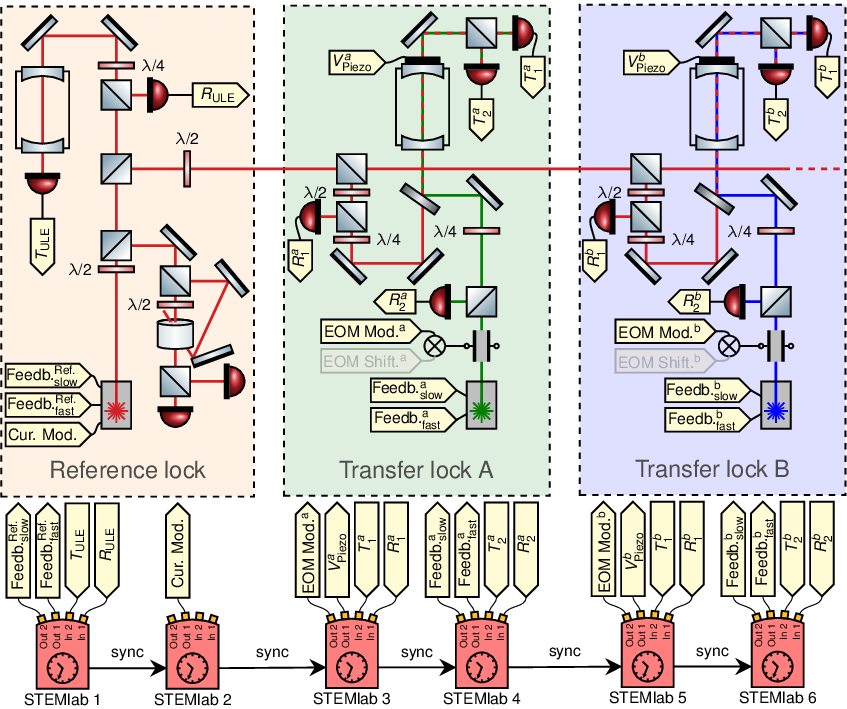}
	\caption{A schematic depiction of the transfer cavity setup as used in our laboratory (detailed
		description in the main text). In the leftmost box a reference laser gets locked to a ULE
		cavity or a DAVLL configuration using PDH.
		Each box to the right indicates an additional laser getting locked to the reference laser
		by means of a transfer cavity (more such systems can be added as required).
		The signals are controlled by means of STEMlab modules which are daisy-chained to share the
		same  clock signal to allow for demodulation of the shared reference oscillation on the
		reference laser.
	}
	\label{fig:setup}
\end{figure}

We show a sketch of our setup in \fref{fig:setup}. Here, we divided it logically into the reference lock and two subsequent transfer locks \enquote{A} and \enquote{B}.
Each transfer lock  can be replicated such that it meets the amounts of locks needed.
The lower part of the figure shows the required electronics, which is based on the use of several STEMlab modules and discussed later.
We discuss the optical setup first.
\par
The reference lock is based on a reference laser which is either stabilized to an ULE cavity or a (rubidium) vapor cell.
A choice between both possibilities is made by using the combination of a half--wave plate \enquote{$\lambda/2$} and a PBS.
In our setup this laser operates at \SI{780}{\nano\m}.
The stabilization to the ULE cavity is accomplished by employing the introduced PDH locking technique.
The laser is coupled into the ULE and the transmission monitored by a photo diode (PD) labeled \enquote{$T_{\text{ULE}}$}.
A quarter waveplate ($\lambda/4$) and another PBS guides the reflected light to another PD labeled \enquote{$R_\text{ULE}$}.
It is possible to use DAVLL as an absolute frequency reference as seen in \fref{fig:PS_DAVLL}, though we currently don't make use of this possibility. Still, this is sketched in \fref{fig:setup}.
\par
Each transfer lock is optically designed the same way.
Our transfer cavities are self-built \cite{Tomschitz.2018}, roughly \SI{15}{\centi\m} in length, and have an FSR of about \SI{900}{\mega\hertz}.
Their mirrors are coated such, that they reflect both the wavelength of the reference laser as well as the wavelength of the target laser.
The reference laser is used to stabilize the length of the transfer cavities.
As this can only be done as soon as the reference laser itself is locked to the ULE, we are not able to scan the laser anymore to observe the ac{PDH} signal.
Thus, we equipped each transfer cavity with a piezo attached to one the mirrors, such that it is possible to modulate the transfer cavity's length.
The piezo's voltage input is labeled as $V_\text{Piezo}$.
We use a dichroic mirror to overlap both beam paths.
A beam splitter (BS) behind the transfer cavity equally splits the passing light between two PDs labeled as $T_1$ and $T_2$, and used for monitoring the transmission.
By using a bandwidth filter the PD either monitors the reference or the target laser.
As the resonance frequency of the transfer cavity is not necessarily equal to the targeted locking wavelength, we guide let the target laser pass an EOM to introduce sidebands for offset locking.
The EOM's modulation input is labeled as \enquote{EOM Mod.}.
We then monitor and demodulate the reflected light of both laser beams labeled as \enquote{$R_1$} and \enquote{$R_2$}, respectively.
\par
The electronic setup is based on several RedPitaya STEMlab 125-14 (RP)s, which are controlled by our slightly modified version of PyRPL \cite{Neuhaus.2017,Neuhaus.2023}.
A single RP board features two inputs and two output ports, and PyRPL implements an arbitrary signal generator (ASG), an in-phase and quadrature (IQ) demodulation, and a PID as well as a scope running all directly on the RP's field--programmable gate array (FPGA).
For the PDH locking technique we need a phase-modulated laser.
We use a first RP to modulate the current of our reference laser, in our case at \SI{15}{\mega\hertz}.
This corresponds to module 2 in \fref{fig:setup}.
On another RP (1 in \fref{fig:setup}) we monitor the transmission and reflection of the ULE.
This RP is also used to generate the PDH error signal out of the monitored reflection.
However, since this is a separate board, the modulation's board clock and demodulation's board clock are not in sync.
As such it is impossible to generate a reliable locking signal, as the signal would drift over time.
We adjusted PyRPL such, that the clock source of the very first RP is daisy-chained to all others by means of SATA--cables.
This enables not only the reliable locking of the reference laser, but also the subsequent locking of the transfer cavities, which is also based on the demodulation of the reference laser's signal.
For any RP controlling the lock of a target laser such a modification is not necessary, as long as only a single board can be used.
Our reference laser is a Toptica DL Pro, which enables us to give both fast (current modulation) and slow (piezo tuning) feedback to our laser using PIDs, as labeled accordingly in \fref{fig:setup}.
For a transfer lock, i.e. a lock stabilizing the length of a transfer cavity, another RP (3 in \fref{fig:setup}) is used, connected to the piezo via an output port, and monitoring the transmission and reflection of the two respective PDs.
Here, we only use a single PID for the stabilization.
For any target laser we add an EOM, which is in our case wired to a Windfreak SynthUSBII as the RF source for offset locking.
Other than that the wiring is identical to that of the reference laser, an exemplary shown with RP 4 in the figure.
Depending on the target laser's type either a single or more PIDs are used.
Next, we will focus on the typical locking procedure.
\subsection{The locking process}
\begin{figure}
	\centering
	\includegraphics{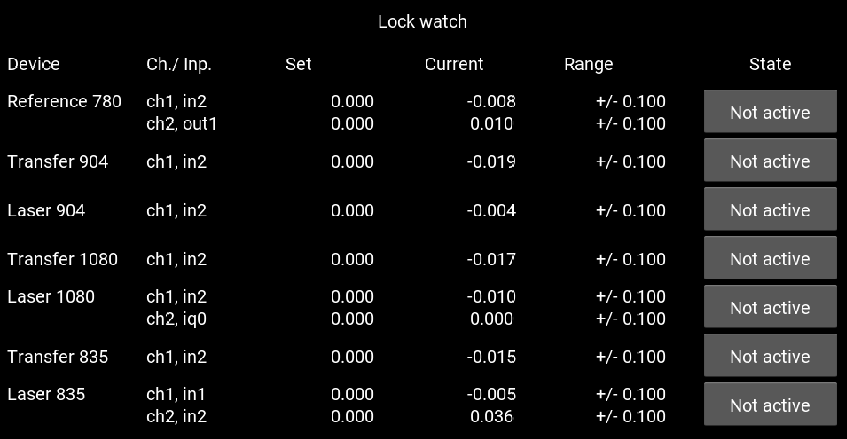}
	\caption{Screenshot of the graphical user interface implemented by us on top of PyRPL showing an overview screen include the lock watches.
	}
	\label{fig:GUI_overview}
\end{figure}

\begin{figure}
	\centering
	\includegraphics{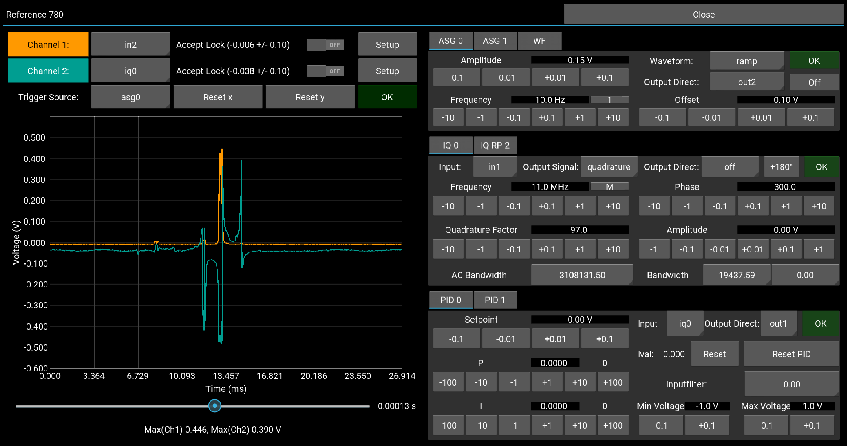}
	\caption{Screenshot of the graphical user interface implemented by us on top of PyRPL showing the reference laser lock screen during the locking sequence.}
	\label{fig:GUI_detail}
\end{figure}

In principle, we could span several UI's provided by PyRPL for each RP.
However, to ease the usability and further implement a \enquote{lock watch}, we wrote a wrapper around PyRPL, which essentially spans a separate PyRPL process per RP, but combines them all in a single interface.
Screenshots can be found in \fref{fig:GUI_overview} and \fref{fig:GUI_detail}.
As any transfer cavity lock relies on the lock of the reference laser, and any target laser lock relies on the lock of the transfer cavity, we always have to lock in order.
\par
We start by locking the reference laser.
At first, we turn on the current modulation of the reference laser by using the ASG of the second RP.
Then we use an ASG of the first RP to apply a ramp to the reference laser's slow feedback port.
On the RP's scope we observe the $00$--mode transmission peaks of the ULE as well as the internal ramp signal, and are able to center such a peak on the ramp by adjusting the ramp's offset.
We continue this process whilst lowering the ramp's amplitude.
When the amplitude is somewhere below \SI{.1}{\volt}, we change the output to only the DC voltage.
Next, we repeat the same process with the fast feedback port and another ASG module of the same RP.
As soon as the signal is sufficiently centered we take a look at the PDH error signal generated by the IQ module of the RP with the input information being the reflected signal as well as the modulation frequency.
Typically the phase is not correct at first and the PDH signal is weak, yet we are able to correct for this from within PyRPL.
Next, we adjust the setpoint of a PID, which gives feedback to the fast feedback port, such, that it is in the midst of this error signal.
We then slightly increase its \enquote{P}--part.
By observing the transmission peak, which should broaden if the phase is correct, we are able to determine if we are \SI{180}{\degree} off or not.
If the phase is correct, we continue locking by disabling the ASG and firstly increasing the \enquote{I}--part and subsequently the \enquote{P}--part of the PID.
Both are tuned such that they are as large as possible without inducing an unstable oscillation on the transmission signal.
If locking was successful, the transmission signal should have turned into a flat line roughly at the maximum of the former transmission peak.
To increase the stability of our lock, we use a second PID which gives feedback to the slow feedback port based on the fluctuations of the first PID.
It uses the output voltage of the first PID as error signal and a setpoint of zero which makes sure that the fast PID has maximal range for rapid adjustments in any direction.
The \enquote{P} and \enquote{I} values are tuned identically to the first PID by slowing increasing their value and observing the onset of unstable oscillations.
If everything went well we are able to turn a \enquote{lock watch}, which simply checks if the mean of the transmission signal stays within a certain range.
\par
In a next step we can lock our first transfer cavity.
Here, an ASG of the respective RP is used to apply a ramp to the piezo.
Again, we observe the transmission peaks of the reference laser, and are able to decrease the scan amplitude while keeping the transmission signal centered.
The IQ module generates the error signal, and the PID then gives feedback to the piezo.
Here, only a single PID is used.
\par
As a final step, we lock our target laser.
In comparison to the lock of the reference laser the only difference is, that we can also adjust the EOM's offset frequency.
For example, if we would like to lock to a certain atomic or molecular transition, we make sure that we have both the cavity signal and the transition signal available, and then tune the frequency such, that the error signal is in line with the transition signal's peak.
We then continue locking with this signal.
\par
\subsection{Characterization}
The performance of the overall system was evaluated by observing both long term drifts of
the locks and the short term linewidth of the target lasers.
Long term stability in our case is primarily limited by disruptive events such as air pressure
fluctuations created by moving doors or strong acoustic noise that couples particularly well
to the transfer cavities.
Whilst the reference laser coupled to a ULE cavity under vacuum often stays locked for
many days or even weeks without interruption, the less shielded environment of transfer cavities
in air tends to produce a loss of lock for some cavity or laser on a daily basis.
We confirm that e.g. a \SI{422}{\nano\meter} ECDL laser locked to the
Rb $5S_{1/2} \to 6P_{1/2}$ transition stays within the natural linewidth during this time.
These disruptive events are heuristically observed by monitoring all control signals and assuming a
loss of lock as soon as an unusually fast change is detected.
Often it is possible to perform automatic relocking by setting the relevant control signals back
to their last recorded locked state.
To demonstrate the short term linewidth, we show data taken using a
\SI{1016}{\nano\meter} ECDL laser locked to the Rb $6P_{1/2} \to 32S_{1/2}$ transition.
\par
This type of characterization can provide a first idea of the overall locking performance, which is
sufficient to verify the suitability of the system for the aforementioned tasks.
A single quantity like the bespoke linewidth in the presence of various noise sources
is only an approximation (although a successful one) as the vague definition
about long-term drifts and short-term linewidths above indicated.
Beyond that, depending on the necessary performance and system characteristics,
a variety of tools are available to describe and measure system noise, long-term locking
stability and frequency resolved laser lineshapes.
In this tutorial we can not cover these topics in detail but just list some common
terminology and references:
The noise performance of all electronic components including a model of the feedback loop
electronics might be used to predict fundamental the limits of a laser lock \cite{Taubman.2013}.
The Allan variance measures how well an oscillator (or more specifically a laser) performs over an
extended period of time as noise continuously averages out \cite{Allan.1966}.
Such measurements, plotted on a logarithmic time axis are shown in many of the cited references
to prove the absence or presence of limiting physical processes over multiple octaves of time
\cite{Hall.2010, Riley.2008}.
In the frequency domain, one usually reasons about the spectral noise density, which can
theoretically be linked to the observed laser linewidth and Allan variance.
Properly measuring actual laser linewidths and lineshapes beyond the directly available,
electrical residual error signals requires an independent optical reference \cite{Hollberg.1990}.
The convenient self-heterodyne linewidth measurement uses a frequency shifted and delayed portion
of the laser itself as a reference but requires careful analysis to avoid especially
when dealing with narrow linewidths where delaying more than the coherence time of the laser
becomes impractical \cite{Huang.2017}.
A heterodyne beat note measurement with a known reference laser is generally preferred but
requires the availability of such a system.
If the beat note between at least two unknown but similar lasers is observed, it is possible to
deconvolute or upper-bound the linewidths of each individual system.
\section{How to fabricate a vapor cell}
The age of modern atomic spectroscopy was initiated by the invention of the low pressure sodium lamp
in 1919, when Arthur H. Compton signed a patent in the USA as an employee of
Westinghouse Electric \cite{Compton.1931}.
This lamp was operated at a pressure of a few \si{\milli\bar} to maintain
enough ionizing collisions for good brightness and featured a comparably narrow spectrum.
As we  know from our work a century later, alkali vapor cells tend to become opaque when operated
at elevated temperature.
Compton also ran into this problem and developed a new type of glass \cite{Compton.1926},
which was a very boron rich variation of the usual borosilicate glass invented by Otto Schott in 1887.
Most of the standard spectroscopy cells, especially reference cells, are made of borosilicate glass to
this day, which run under brand names like Duran, Pyrex, Borosil, Borofloat, and many others.
\par
Yet there are many other materials beside borosilicate glass, which can be used to make vapor cells.
These are supplemented by a plethora of fabrication processes beyond the traditional glassblowing.
In the following subsections we will introduce to the most common materials and methods
and contextualize their use within the field of vapor cell fabrication.
\par
A more extensive review for the specific case of microelectromechanical system (MEMS) vapor cells can be found in the PhD
thesis of Karlen \cite{Karlen.2017} and the article by Knapkiewicz \cite{Knapkiewicz.2018}.
\subsection{Design considerations}
In the first part of this tutorial we introduced vapor cells in the form of reference cells as a
tool for locking lasers.
In a broader sense, we want to cover any spectroscopy cell, i.e. defined volume, which holds atoms
or molecules in gaseous state to be probed by lasers, electric and magnetic fields.
Depending on the temperature, we speak of a vapor cell as long as the temperature of the medium
stays below the critical temperature and thus a portion stays liquid or solid.
As soon as this critical temperature is surpassed, the same cell can become a gas cell in which all
particles are gaseous.
\par
We can also differentiate such cells depending on the idealized thermodynamic flow of particles:
\begin{itemize}
	\item Open cells are connected to their environment, e.g. through a series of pipes, which
	allows exchange of particles with an external reservoir.
	This typically requires some form of
	active pumping or evacuation to uphold specific flow or pressure conditions.
	\item Sealed (or closed) cells are disconnected from the environment, thus allowing no further
	particle exchange.
	As we will discuss below, actual implementations might still permeate and desorb atoms or include
	getter materials to ensure constant conditions.
\end{itemize}
\par
There are many applications beyond the laser locking reference, which see the use of spectroscopic
cells such as sensing, atomic clocks, research on electromagnetic interactions
(EIT, photon storage, etc.), and gas analysis.
\par
Thus, depending on the intended application, very different requirements can arise.
\Tref{tab:fabrication_criteria} lists many of these criteria which can serve as a
starting point during the planning phase of a vapor cell design.
In the following sections we will focus on common materials and fabrication techniques.
\begin{table}
	\centering
	\caption{\label{tab:fabrication_criteria}
		A summary of aspects and constraints to consider during the planning and
		fabrication of a vapor cell design sorted into sections.
		The left column contains key criteria while the right column provides
		additional context and specific circumstances under which the criteria
		become relevant.
	}
	\footnotesize
	\begin{tabular}{@{}ll}
		\br
		\centre{2}{Chemical Compatibility and Atomic Medium}\\
		\mr
		Atomic medium&\textbullet\ Required amount and purity, method of filling\\
		&\textbullet\ Cell materials must be chemically compatible\\
		Additional internal media&\textbullet\ Buffer gases, getter materials, chemical reactants\\
		Leak Tightness&\\
		Lifetime / Aging&\textbullet\ Chemical degradation, deposition, leakage, and contamination \\
		&\textbullet\ Reaction with and diffusion into cell surface\\
		\mr
		\centre{2}{Optical, Electronic, and Magnetic Interface}\\
		\mr
		Transmitted wavelengths&\textbullet\ Suitable material, polished surfaces, unobstructed optical path\\
		&\textbullet\ Might require anti reflection coating, Brewster angled windows\\
		Maximum optical power&\textbullet\ Damage threshold and optically induced chemical reactions\\
		Additional optical functions&\textbullet\ Photonic structures, waveguides, optical fibers\\
		&\textbullet\ Internal deflection (mirror surfaces, prisms, gratings)\\
		Electronic contacts&\textbullet\ Applying AC/DC fields, electrical heating, electronic readout\\
		Magnetic shielding&\textbullet\ Typically non magnetic materials to avoid frequency shifts\\
		\mr
		\centre{2}{Environmental Conditions}\\
		\mr
		Temperature and Humidity&\textbullet\ During operation and fabrication (typically non-condensing)\\
		Pressure&\textbullet\ Typically ambient pressure, might be used for length tuning \\
		Surrounding chemistry&\textbullet\ Protective atmosphere, laboratory, or rough environment?\\
		\mr
		\centre{2}{Physical Parameters}\\
		\mr
		Shape, Volume, Mass&\textbullet\ Size of enclosures, ovens, required optical access\\
		Mechanical Strength&\textbullet\ External mechanical and thermal loads\\
		\mr
		\centre{2}{Economic Constraints}\\
		\mr
		Manufacturability&\textbullet\ Single prototypes, many different designs, or industrial scale\\
		&\textbullet\ Expected yield, availability, and time scale of fabrication\\
		Cost&\\
		\br
	\end{tabular}
\end{table}
\subsection{Bulk materials for vapor cells}
The most common material found in vapor cell is glass, due to its transparency, vacuum tightness and -- depending
on the type -- chemical compatibility with alkali vapors.
Its most striking feature is the ability to produce high quality, vacuum tight joins by means of conventionally glassblowing.
There are however a range of materials beyond glass that can be used such as silicon, e.g.
in parts of the vapor cell that don't transmit visible light.
\par
Most glassblown cells are made of hard glass with a high softening point and low
thermal expansion coefficients below \SI{60e-7}{\per\kelvin}
like borosilicate glass or fused silica (amorphous quartz) \cite{Jong.2000}.
Fused silica (SiO$_2$) typically has better optical properties than borosilicate and exists
in a large variety of purities.
Well known manufacturers are Heraeus (Herasil, Suprasil, Homosil, Infrasil) and Corning
(HPFS High Purity Fused Silica 7979, 7980 and 8655) among many others.
Borosilicate only contains \SIrange{70}{80}{\percent} SiO$_2$. Other ingredients are B$_2$O$_3$, Na$_2$O, K$_2$O, Al$_2$O$_3$, and other oxides. The alkali contribution is particularly important for anodic bonding as described below. The main manufacturers are Corning (Pyrex) and DWK Life Science, formerly Schott (Duran). Typical properties of fused silica and borosilicate glass are summarized in table \ref{glass_materials}.
\par
Soft glasses such as soda-lime glass have an even higher alkali fraction and are typically of lower optical quality.
Their lower softening point reduces thermal and mechanical shock resistance.
Many optical glasses like crown glass and flint glass are not well suited for glassblowing.
\begin{table}
	\caption{\label{glass_materials}Typical properties of glass materials used for the fabrication of vapor cells.}
	\footnotesize
	\begin{tabular*}{\textwidth}{@{}l*{15}{@{\extracolsep{0pt plus
						12pt}}l}}
		\br
		&Fused Silica& Borosilicate \\
		\mr
		Softening point & $\approx\SI{1600}{\degreeCelsius}$	 & $\approx\SI{1200}{\degreeCelsius}$		\\
		Strain (annealing) temperature & $\approx\SI{1050}{\degreeCelsius}$ & $\approx\SI{540}{\degreeCelsius}$ \\
		Thermal expansion coefficient & $\approx\SIrange{5e-7}{6e-7}{\per\kelvin}$ & $\approx\SI{3.3e-7}{\per\kelvin}$ \\
		Transparency & $\approx\SIrange{185}{3500}{\nano\meter}$  & $\approx\SIrange{350}{2000}{\nano\meter}$ \\
		\br
	\end{tabular*}
\end{table}
\par
Beyond glass, the fabrication and joining of the vapor cell typically becomes more complicated as we will discuss in the sections below.
Sapphire (Al$_2$O$_3$) comes to mind as a well known anti-diffusion
coating in its amorphous form (called alumina).
Bulk sapphire alkali vapor cells have been successfully made and are used for research \cite{Sarkisyan.2001}
involving high temperature at which glass-made alkali vapor cells already exhibit
excessive diffusion \cite{Peyrot.2018}.
\par
Silicon (Si, and to a lesser extend also related materials like silicon nitride Si$_3$N$_4$ and silicon carbide SiC) forms the basis of many contemporary activities in the fields of nanophotonics
and MEMS vapor cells as it allows to integrate functional devices into vapor cells.
The advanced capabilities and methods of the semiconductor industry allow for complex nanostructuring and integration which lead to the miniaturization of vapor cell based devices such as atomic clocks or magnetometers.
Silicon itself is compatible with most alkali metals, partly transparent for infrared light, and vacuum tight in its crystalline form.
\par
Other notable material candidates for vapor cell fabrication include crystalline quartz, ceramic materials, metals, and potentially even plastics.
Metals play a particularly important role in the context of anodic bonding as discussed in the next section and form the basis of most open gas cell vacuum systems.
\par
Due to the strong influence of buffer gas pressures and amount of the atomic medium
in the context of frequency reference vapor cells, much attention has been spent on the
interaction between these materials and alkali vapors \cite{Karlen.2017}.
It turns out that both fused silica, and borosilicate glass have a good resistance against
alkali vapors, while soft glasses tend to be easily reduced.
The reactivity of lithium in combination with the rather high melting point of \SI{180.5}{\degreeCelsius} stands out here and does not allow for a stable operation of a vapor cell as we discuss here. Therefore, the vast majority of cells are filled with rubidium and cesium, and in some occasions with sodium and potassium. If one wants to work with lithium vapors it is indispensable to return to the very well established techniques of discharge cells \cite{Slabinski.1971} and heat pipe ovens \cite{Vidal.1969}, which are used up to date.
Compatibility of metals with alkalis has been made in terms of corrosion and solubility of he alkali in the metal \cite{Karlen.2017}.
\par
It is important to know that one cannot easily combine each pair of materials due to their different thermal expansion coefficients.
A temperature change of a hybrid cell of e.g. fused silica and borosilicate glass leads to immediate destruction from internal stress already in the production state.
If multiple materials must be combined, they either need to have matching coefficients of thermal expansion or a smooth transition from one to the other must be created.
In the case of the fused silica to borosilicate glass transition, adding layers with different alkali concentration transitions the thermal expansion coefficient smoothly as it is done for glass metal transitions in vacuum technology.
\par
\subsection{Functional internal coatings and materials}
While the outer walls of a vapor cell primarily need to provide a vacuum tight enclosure for the atomic or molecular medium, a wider range of functional materials and parts can be integrated inside those boundaries.
These include e.g. optical fibers, mirrors, surface coatings, getter materials, and buffer gases.
The available materials are still restricted by the chemical comparability but might also protect the cell walls.
\par
Anti-relaxation coatings have been used as early as the late 1950s \cite{Franzen.1959} and are used up to this day to increase the spin-polarization lifetime in atomic media \cite{Chi.2020}.
The most common material in this context is paraffin with a number of alternatives to be used e.g. at elevated temperatures.
\par
Anti-diffusion coatings decrease the diffusion rate of the atomic medium into the vapor cell walls.
We use dense alumina (Al$_2$O$_3$) layers made from atomic layer deposition (ALD) of about \SI{10}{\nano\meter} thickness to prevent cesium and rubidium from attacking glass surfaces made of fused silica or borosilicate glass \cite{Woetzel.2013}.
\par
Vapor cells with thin-film anti-reflection coatings are commercially available.
In a similar manner it is possible to create reflective thin-film reflectors inside vapor cells \cite{Perez.2009}.
These coatings should only employ materials that inherently resist the atomic media or might be closed with a anti-diffusion layer.
\par
Unwanted gases that reside inside the vapor cell can lead to adverse effects such as pressure broadening or decreased spin-relaxation lifetimes.
While this can be actively minimized by a vacuum bakeout and sufficient pumping during the filling stage, a closed vapor will eventually leak, outgas, or form gases from internal chemical processes.
This is counteracted by the inclusion of so-called getter materials, which react with the vapor cell atmosphere and binds any unwanted trace gases such as H$_2$O or O$_2$ \cite{Giorgi.1985}.
\par
Finally, buffer gases are purposely added to vapor cells for various reasons.
In alkali vapor frequency standards, buffer gases are added to narrow hyperfine transitions below the normal Doppler width by increasing the effective lifetime inside the laser beam \cite{Brandt.1997} (similar to anti-relaxation coatings).
\subsection{Available fabrication processes for sealed vapor cells}
All aforementioned materials can be processed using a wide range of techniques including conventional milling, fine optical processing, micro-structuring, and scientific glassblowing.
The eventual challenge is the production of a durable and vacuum tight seal.
We can distinguish between two main methods for vacuum tight cell joining: scientific glassblowing and mechanic-chemical bonding.
One could potentially even produce a sealed metal based vapor cell using conventional vacuum flanges and components.
Maintaining a low enough pressure will however typically require active pumping in such cases.
\subsubsection{Cells joined by scientific glassblowing}
\begin{figure}
	\centering
	\includegraphics{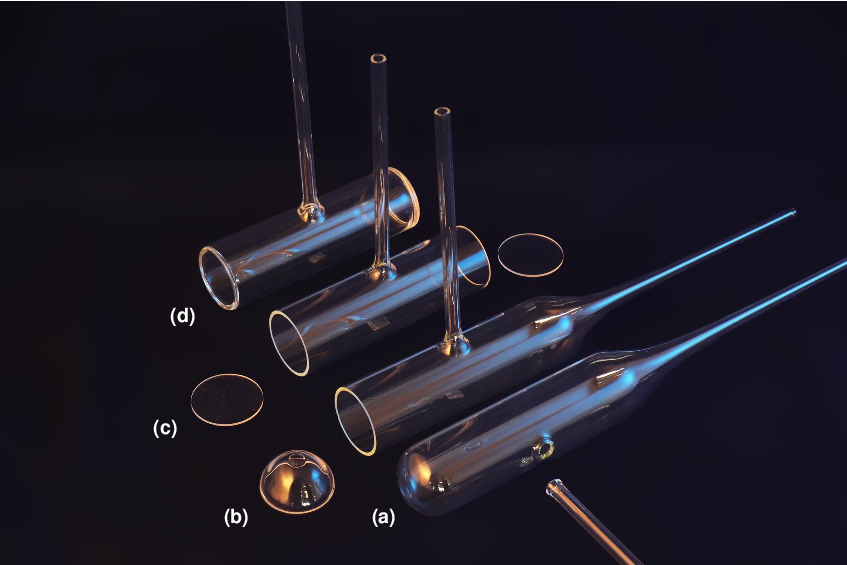}
	\caption{Steps to make a reference cell using conventional glassblowing techniques (details in the text).}
	\label{fig:reference_cell_making}
\end{figure}

\begin{figure}
	\centering
	\includegraphics{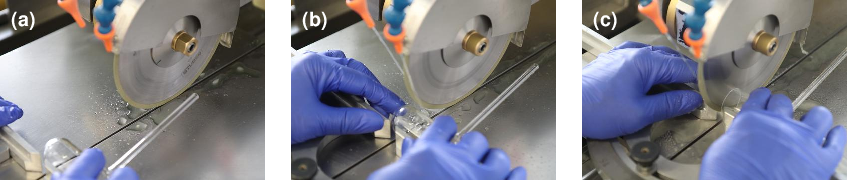}
	\caption{Cutting the glass tube to the desired length (see \fref{fig:reference_cell_making}(b)).}
	\label{fig:cutting_tube_tablesaw}
\end{figure}

\Fref{fig:reference_cell_making} outline a typical progression during the manufacturing of a reference vapor cell.
While these steps are usually performed by a skilled scientific glassblower \cite{Moore.2009}, we will highlight some important steps to consider during a scientific planning phase.
In the first step a borosilicate glass tube of \SI{30}{\milli\meter} diameter with \SI{1.5}{\milli\meter} wall thickness gets cut to the desired length by developing pulling points on both ends.
The left side in \fref{fig:reference_cell_making}(a) is fully formed into a test-tube like end, while the right side is kept as a holder to handle the part in the following steps.
A T-seal joint towards a tube of smaller diameter is prepared which is going to serve as access point for evacuation and filling.
In the next step, the ends are cut off using a diamond glass cutting disc as shown in \fref{fig:cutting_tube_tablesaw} under continuous water cooling.
\par
The windows which get prepared in \fref{fig:reference_cell_making}(c) are polished to the desired optical quality and have approximately the same thickness as the tube's walls.
The rough surfaces on the tube's face and the window's edges get fire-polished to prevent cracking and ensure a leak-free joint.
The windows and the tube get joined using a glassblowers lathe in which the tube is held in a 3-jaw chuck on one end, while the window is hold using a vacuum chuck.
\par
While glass is only a weak heat conductor and allows e.g. to hold the part during the fabrication, the required melting temperatures typically prevent the presence of most coatings during the joining process.
A slight variation of purely glassblown cells is the method of glass frit bonding \cite{Ramm.2012}.
Here, the individual parts are connected by a lower melting point glass powder or paste as an intermediate bonding layer.
The assembly with the frit applied to the interfaces is then placed in an oven at a set temperature below \SI{450}{\degreeCelsius} and pressure where the frit locally melts without the remaining cell.
Frits of suitable thermal expansion exist for almost any material, especially for borosilicate, quartz, and even sapphire.
Most commercially available, macroscopic reference alkali vapor cells are either made by scientific glassblowing or frit bonding.
\subsubsection{Anodic bonded cells}
Anodic bonding \cite{Wallis.1969} is one of the most common methods to produce MEMS type cells, especially on an industrial scale.
Typically two borosilicate windows are anodically bonded to a silicon spacer.
By applying heat (up to a few \SI{100}{\degreeCelsius}) and high voltage (up to a few \SI{100}{\volt}) across the stack, ions in the borosilicate (mostly sodium) start to diffuse away from the silicon while oxygen diffuses towards the silicon.
The negatively charged oxygen ions form a covalent bond with the silicon.
Its bond strength can be improved by adding a thin layer of aluminum, which leads to a connecting Al$_2$O$_3$ layer.
The first cell based on two borosilicate windows bonded to a silicon spacer has been realized in 2004 \cite{Liew.2004}.
Most commercial magnetometers and atom clocks us cells made by anodic bonding.
For other applications is also possible integrate vacuum compatible electric feedthroughs \cite{Daschner.2012}, e.g. to apply electric fields in a controlled manner or for direct current readout of highly excited Rydberg states \cite{Barredo.2013}.
\subsubsection{Other cell methods to join alkali vapor cells}
Besides glassblown cells and cells made by anodic bonding many other methods have been developed to fabricate a vacuum tight, long lived spectroscopy cells for alkali atoms.
The most simple approach is to use a vacuum tight glue.
In most cases this is not a suitable solutions as these cells cannot be operated at elevated temperatures, due to the differing thermal expansion coefficients, and a reduced lifetime of a few weeks to months.
It is not clear, whether the cells develop leaks, the glue outgases or chemical reactions of the alkali atoms reduce the lifetime.
A typical observation is, that a background pressure develops over time and pressure broadening prohibits spectroscopy at some point.
This process is proportional with temperature.
Cells sealed with vacuum epoxy like Torr Seal can last many months when operated at room temperature.
One exception to this maxim are the cells produced in Armenia by the group of D. Sarkisyan.
Their sapphire cells are joined using a mineralic high-temperature solder \cite{Sarkisyan.1989} adapted form the fabrication of sodium vapor lamps which belong to the class of eutectic \SI{49}{\percent} Al$_2$O$_3$, \SI{45}{\percent} CaO, \SI{6}{\percent} MgO compositions with added ZrO$_2$ \cite{McVey.1979,Gavrilov.2019}.
\par
A less common method resulting in robust vacuum tight vapor cells is thermocompression \cite{Karlen.2020}.
Here, a copper layer is added to the two pieces to be connected and then pressed onto each other at elevated temperature.
In \cite{Karlen.2020} they used \SI{500}{\nano\meter} of copper which was fused at \SIrange{350}{400}{\degreeCelsius} and a pressure of \SI{14}{\kilo\newton}.
The main advantage of this method is that other materials beyond borosilicate can be used for the cell as the mobility of ions in borosilicate (anodic bonding) is not necessary.
For cells with very small volumes this method has the additional benefit of reduced residual outgasing compared to anodic bonding.
\par
Although sealing vapor cells by local heating with a laser beam seems to be an obvious method, not too many attempts along this line have been carried out.
Researchers mostly used a CO$_2$ laser to apply localized heating at a certain position to perform effectively glassblowing on a microscopic scale \cite{Knappe.2003,Losev.2015}.
The possibility to apply the laser beam only at a specific spot, leaving the rest of the cell at low temperature, has been exploited to make a MEMS type cell with a local breakseal \cite{Maurice.2022}.
\par
The last glass joining method mentioned here is direct bonding by optical contacting and by chemical activation.
Both methods do not require elevated temperatures but highly polished surfaces \cite{Peyrot.2019b,Cutler.2020}.
\par
Applications requiring pressures higher than atmospheric conditions \cite{Larson.1991} can be made using steel chambers in combination with metal flanges joined by active soldering \cite{Ockenfels.2021}.
\subsection{Filling of alkali vapor cells}
The filling of sealed vapor cells with highly reactive alkalis can be achieved in different manners.
The most common method relies on pure alkali samples, which are transferred from a sealed glass ampule into the cell (some good illustrations are found in \cite{Missout.1975}, \cite{Knapkiewicz.2018}, and \cite{Maurice.2022}).
To do so, the cell to be filled is connected by glassblowing to a breakseal ampule via a manifold.
This manifold is connected to a turbo pump using e.g. a glass Kleinflansch that can be connected in the usual way to a metal Kleinflansch.
The whole assembly and especially the parts that will be in contact with the alkali should then be baked for a couple of days above \SI{100}{\degreeCelsius}.
The pressure after bake-out at room temperature should be on the order of \SI{10e-7}{\milli\bar} or better.
To open the ampule, we use a little metal ball which is smashed on the glass nozzle of the breakseal of the ampule with the help of a magnet from the outside.
The alkali metal is then driven into the cell by gently heating the metal with a heat gun and shaking the liquid alklai metal into the cell.
Finally the cell is pinched off while the assemly is still attached to the vacuum pump.
\par
This basic process leads to a macroscopic droplet of the alkali deposited in the vapor cell, which can be undesirable due to the application, the price of e.g. isotopically enriched rubidium, or the purity of the available material.
It is instead possible to distill the alkali from the breakseal ampule to the vapor cell by heating the manifold and creating a condensation spot in the vapor cell.
In this case we close the connection to the vacuum pump during distillation e.g. with a metal ball \cite{Missout.1975} to prevent the loss of the material.
We perform distillation of rubidium at slightly above \SI{100}{\degreeCelsius} (having the cold spot at room temperature) which allows us to control the transferred amount by observing the formation of a thin-film rubidium layer over the course of several minutes to hours.
Distillation of cesium is possible even at room temperature by introducing a cold spot with ice water.
\par
Any buffer gases can be introduced into the vapor cell at this stage by carefully opening a flow from a gas bottle or gas mixing unit.
In this case, a two step sealing process might be required to ensure predictable pressure conditions \cite{Missout.1975}:
First, the alkali is brought only close to the final location and then removed from the manifold togher with the cell as usual.
Then, in the already enclosed system, the alkali is moved or distilled into the actual vapor cell.
Such a two step process is also advantageous if the manifold is impracticably large or fragile.
\par
For small scale cells, as used for magentometry and atom clocks, this filling method is not very well controlled and many different methods have been reported \cite{Karlen.2017,Knapkiewicz.2018}.
Direct transfer or pipetting of molten alkalis requires an artificial, protective, non-reacting atmosphere.
Knappe et. al. \cite{Liew.2004} used a chemical reaction between barium azide and alkali chloride which produces at elevated temperature elemental alkali and nitrogen.
In a similar manner one can decompose RbN$_3$ with the help of UV light \cite{Karlen.2017b}.
With both methods it is possible to adjust a specific partial pressure of nitrogen.
\par
For some applications it is necessary to avoid any buffer gas to reduce collisional broadening and excited state quenching. In this case one can use small samples of alkalis sealed in a wax layer to avoid oxidation \cite{Radhakrishnan.2005,Wang.2022b}.
\section*{Acknowledgments}
The authors thank Frank Schreiber for the preparation of the glass vapor cells.
We also thank Patrick Kaspar for support during the setup phase and Philipp Neufeld for contributions to the code base.
\par
This letter is supported by the Deutsche Forschungsgemeinschaft (DFG)
via Grant LO 1657/7-1 under DFG SPP 1929 GiRyd as well as Grant DFG 431314977/GRK2642 (Research Training Group:
``Towards Graduate Experts in Photonic Quantum Technologies'').
This project has also received funding from the European Union's Horizon 2020 research and innovation program under Grant Agreement No.~820393 (macQsimal).
\section*{List of Acronyms}
\begin{description}
\item[ALD] atomic layer deposition
\item[AOM] acousto-optic modulator
\item[ASG] arbitrary signal generator
\item[BS] beam splitter
\item[CTE] coefficient of thermal expansion
\item[DAVLL] dichroic atomic vapor laser lock
\item[DBR] distributed Bragg resonator laser
\item[DFB] distributed-feedback diode laser
\item[ECDL] external-cavity diode laser, sometimes also called extended-cavity diode laser,
\item[EIT] electromagnetically induced transparency
\item[EOM] electro-optic modulator
\item[FM] frequency modulation
\item[FMS] frequency modulation spectroscopy
\item[FPGA] field--programmable gate array
\item[FSR] free spectral range
\item[FVC] frequency to voltage converter
\item[FWHM] full width at half maximum
\item[IQ] in-phase and quadrature
\item[LO] local oscillator
\item[MEMS] microelectromechanical system
\item[MTS] modulation transfer spectroscopy
\item[OFC] optical frequency comb
\item[OFLL] optical frequency lock loop
\item[OPLL] optical phase lock loop
\item[PBS] polarizing beam splitter
\item[PD] photo diode
\item[PDH] Pound-Drever-Hall
\item[PFD] phase frequency detector
\item[PID] proportional-integral-derivative
\item[PM] phase modulation
\item[PSD] power spectral density
\item[RAM] residual amplitude modulation
\item[RF] radio frequency
\item[RP] RedPitaya STEMlab 125-14
\item[SAS] saturated absorption spectroscopy
\item[SHG] second harmonic generation
\item[SNR] signal-to-noise ratio
\item[TiSa] titanium-sapphire laser
\item[ULE] ultralow-expansion
\item[VCSEL] vertical-cavity surface-emitting laser
\item[VECSEL] vertical-external-cavity surface-emitting laser
\item[WM] wavelength modulation
\item[WMS] wavelength modulation spectroscopy
\end{description}
\section*{Data availability statement}
All data that support the findings of this study are included within the article.
\clearpage
\bibliographystyle{unsrtnat}
\apptocmd{\sloppy}{\hbadness 10000\relax}{}{}
\small
\bibliography{bibliography}
\end{document}